\documentclass[twocolumn]{aastex631}
\usepackage{amsmath}

\newcommand{\apc}[1]{\textcolor{black}{#1}}

\newcommand{\coco}{COCO}
\newcommand{\updated}[1]{\textcolor{black}{#1}}

\newcommand{\reply}[1]{#1}

\DeclareRobustCommand{\kms}{\mathrm{km\,s^{-1}}}
\DeclareRobustCommand{\msol}{\mathrm{M_{\odot}}}

\begin{document}

\title[Ghostly Galaxies]{Ghostly galaxies: accretion-dominated stellar systems in low-mass dark matter halos}

\author[0009-0007-3700-9200]{Chung-Wen Wang}
\affiliation{Institute of Astronomy and Department of Physics, National Tsing Hua University, Hsinchu 30013, Taiwan}
\altaffiliation{chungwencw234@gmail.com}

\author[0000-0001-8274-158X]{Andrew P. Cooper}
\affiliation{Institute of Astronomy and Department of Physics, National Tsing Hua University, Hsinchu 30013, Taiwan}
\affiliation{Center for Informatics and Computation in Astronomy, National Tsing Hua University, Hsinchu 30013, Taiwan}
\altaffiliation{apcooper@gapp.nthu.edu.tw}

\author[0000-0002-0974-5266]{Sownak Bose}
\affiliation{Institute for Computational Cosmology, Department of Physcis, Durham University, Durham DH1 3LE, UK}

\author[0000-0002-2338-716X]{Carlos S. Frenk}
\affiliation{Institute for Computational Cosmology, Department of Physcis, Durham University, Durham DH1 3LE, UK}

\author[0000-0003-4634-4442]{Wojciech A. Hellwing}
\affiliation{Center for Theoretical Physics, Polish Academy of Sciences, Aleja Lotników 32/46, 02-668 Warsaw, Poland}



\begin{abstract}
Wide-area deep imaging surveys have discovered large numbers of extremely low surface brightness (LSB) dwarf galaxies, which challenge galaxy formation theory and, potentially, offer new constraints on the nature of dark matter. Here we discuss one as-yet-unexplored formation mechanism that may account for a fraction of LSB dwarfs. We call this the `ghost galaxy' scenario. In this scenario, inefficient radiative cooling prevents star formation in the `main branch' of the merger tree of a low-mass dark matter halo, such that almost all its stellar mass is acquired through mergers with less massive (but nevertheless star-forming) progenitors. Present-day systems formed in this way would be `ghostly' isolated stellar halos with no central galaxy. We use merger trees based on the extended Press--Schechter formalism and the Copernicus Complexio cosmological $N$-body simulation to demonstrate that mass assembly histories of this kind can occur for low-mass halos in $\Lambda$CDM, but they are rare. They are most probable in isolated halos of present-day mass $\sim4\times10^{9}\,\mathrm{M_{\odot}}$, occurring for $\sim5\%$ of all halos of that mass under standard assumptions about the timing and effect of cosmic reionization. The stellar masses of star-forming progenitors in these systems are highly uncertain; abundance-matching arguments imply a bimodal present-day mass function having a brighter population (median $M_{\star} \sim3\times10^{6} \mathrm{M_{\odot}}$) consistent with the tail of the observed luminosity function of ultradiffuse galaxies. This suggests that observable analogs of these systems may await discovery. We find that a stronger ionizing background (globally or locally) produces brighter and more extended ghost galaxies. 
\end{abstract}



\section{Introduction}

Deep imaging observations in the Local Group, around other nearby galaxies, and in galaxy clusters have demonstrated that a large population of very faint galaxies exists below the surface brightness detection limit of current wide-area surveys \citep[e.g.][]{Trentham:2001aa}. These galaxies can provide a low-redshift probe of early galaxy formation, cosmic reionization, and the nature of dark matter. Comparisons of low surface brightness (LSB) dwarfs with theoretical predictions have concentrated on the satellites of Milky Way--like galaxies. With deeper all-sky surveys, improved redshift-independent distance estimates and higher-resolution cosmological volume simulations, it will soon be possible to study this important population in the field, over a much larger volume, and to make more robust statistical comparisons with models.

Observations of so-called ultradiffuse galaxies \citep[UDGs; e.g.][]{van_Dokkum_2015,Koda_2015_UDGs, Torrealba:2016ab,Torrealba_2019_Antlia2} have revived interest in the properties and origins of the LSB dwarf population \reply{\citep[for a recent summary, see][]{zaritsky:2023_smudges}}. It is not clear that the size and surface brightness criteria used to define UDGs\footnote{There is no consensus definition of a UDG; a common choice in the literature is the combination of low surface brightness $\langle \mu \rangle_{\mathrm{eff}} > 24\,\mathrm{mag\,arcsec^{-1}}$ and large physical size $r_{\mathrm{eff}} > 1.5\,\mathrm{kpc}$.} pick out a distinct population of objects that form in a different way from other LSB dwarfs \citep[e.g.][]{Van-Nest:2022wg}. Many formation scenarios for LSB dwarfs have been explored with the motivation of explaining the \reply{UDGs}. These range from the relatively unremarkable high angular momentum tail of standard $\Lambda$CDM galaxy formation \citep{Amorisco:2016aa} to astrophysical processes that may only be relevant at very small scales \citep{jiang2019}. Since many diffuse galaxies have been discovered in clusters \citep[e.g.][]{van-der-Burg:2017aa}, scenarios involving tidal interactions and other effects of dense environments have been considered in most detail \textbf{\citep[e.g.][]{Jones:2021wy, fielder2023}}, although extremely low surface brightness systems also exist in the field \citep{Barbosa:2020vq,Sales:2020vg}. The relative contributions of these different processes to the observed LSB dwarf population remains to be determined.

In this paper we describe a straightforward mechanism that could give rise to a fraction of the LSB field population. In brief, we suggest that some diffuse galaxies may result from the tidal disruption of one or more satellites in a dark matter halo that does not form a central galaxy. We provide a fuller explanation of this idea in the next section. We call systems formed in this way `ghost galaxies'; effectively they are galaxies in which all the stellar mass is associated with an accreted stellar halo component\footnote{The name was inspired by \citet{Lynden-Bell:1995vu}.}. Compared to typical galaxies of similar stellar mass, such galaxies would naturally be more extended and reside in more massive halos  (subject to caveats that we explore below). As we demonstrate, predictions for the luminosity function and halo mass distribution of such galaxies depend strongly on the degree of heating of the intergalactic medium (IGM) by the cosmic UV background. Stronger heating, somewhat counterintuitively, results in more (but fainter) systems of this type, occupying a wider range of halo mass. 

\reply{\citet{peng2016} proposed a scenario in which UDGs form as pure stellar halos, based on the apparently high specific frequency of globular clusters in the Coma UDG Dragonfly 17. This idea is has since been referred to as the `failed galaxy' scenario \citep[see e.g.,][]{pandya2018,fielder2023}. There is only a subtle difference between this and the ghost galaxy concept. In \citet{peng2016} and the subsequent literature, the absence of concentrated star formation is attributed to a feedback mechanism (not specified in detail) acting on the gas in the `failed' galaxy itself, perhaps related to the process of globular cluster formation (implicitly, in the same galaxy). We argue that the hierarchical assembly of halos subject to reionization provides a more straightforward and natural means of creating stellar halos (and globular cluster populations) without proportionally massive central galaxies. We note that \citet{zaritsky:2023_smudges} find evidence that UDGs fall systematically below the relation between stellar mass and halo mass defined by higher surface brightness dwarf galaxies, qualitatively consistent with this scenario, although (as we discuss below) the stellar masses of the UDGs in their sample are substantially higher than those we predict for ghost galaxies.}

To our knowledge, this scenario has not yet been considered explicitly in the \reply{theoretical} literature, although it is a corollary of other well-known aspects of dwarf galaxy formation. It is effectively inevitable in the $\Lambda$CDM model. It does not involve any new theoretical concepts, beyond those already known to be essential to the current understanding of galaxy formation in low-mass dark matter halos. 

Given the emphasis on UDGs in the recent literature, we feel that it is important to emphasize the following two points. First, we do not argue that the ghost galaxy scenario is responsible for all LSB dwarf galaxies. Indeed, we demonstrate that this cannot be the case. Second, we do not argue that it produces all (or even any) of the known objects classified as UDGs. Instead, our aim is only to estimate how common these ghostly galaxies are and whether they are likely to be observable, under some simple but plausible assumptions about their likely stellar masses. \reply{We focus on the field, although in principle ghost galaxies could also occur in clusters, and may have different properties and abundance in dense environments. We discuss dense environments further in our conclusions.}



We proceed as follows. In section~\ref{sec:ghost_idea} we elaborate on the concept of ghostly galaxies. In section~\ref{sec:method} we describe our merger-tree-based methods to quantify the probability of ghost galaxy formation in halos of different masses. Section 3 is the result and comparison of different methods. We summarize our findings in section~\ref{sec:conclusion}. Throughout, for consistency with the Copernicus Complexio (\coco{}) $N$-body simulation, we use the WMAP7 cosmological parameters with $h_{0}=0.704, \Omega_{0}=0.272, \Omega_{b}=0.04455, n_{s}=0.967$, and $\sigma_{8}=0.81$.

\section{The origin of ghostly galaxies}
\label{sec:ghost_idea}

\subsection{Limits on low-mass galaxy formation}

A cornerstone of galaxy formation theory in CDM cosmogonies is that galaxies cannot form through dissipative collapse in dark matter halos with present-day virial mass much less than $\sim10^{10}\,\mathrm{M_{\odot}}$. This idea underpins the concept of a ``ghost galaxy,'' which we introduce in the next subsection. We first give a brief recap of the two fundamental processes that determine the limiting halo mass. For a more complete review, we refer the reader to \citet{Benitez_Llambay_2020}.

\apc{Star formation requires a reservoir of dense, cold (neutral) baryons to accumulate in dark matter halos. This accumulation is expected to follow from the radiative cooling and subsequent inflow of a quasi-hydrostatic atmosphere in virial equilibrium with the dark matter \citep{White:1978aa,White_Frenk_1991}. The ability of a gravitational potential to confine gas from the IGM with an ambient temperature $T_\mathrm{IGM}$ can be expressed as a threshold in virial temperature, $T_{\mathrm{vir}}$; halos with $T_{\mathrm{vir}} < T_\mathrm{IGM}$ cannot accumulate a virialized overdensity of baryons (baryons are said to ``evaporate'' from such halos). In the early Universe, where the ambient IGM is neutral, $T_\mathrm{IGM}$ is low. However, in halos that can accumulate baryons, radiative cooling can only proceed efficiently if the equilibrium temperature of those baryons (i.e. the virial temperature of the halo, $T_{\mathrm{vir}}$) is high enough to maintain them in collisional ionization equilibrium at their equilibrium density  \citep{White:1978aa}. If this is not the case, the gas will remain stable against radiative cooling and condensation. A critical virial temperature, $T_{\mathrm{vir,c}}\sim10^{4}\,\mathrm{K}$, can be associated with this limit, corresponding to the temperature at which atomic hydrogen is ionized in collisional equilibrium. This ``atomic hydrogen cooling floor'' puts a stringent limit on the population of dark matter halos that can support galaxy formation in the early Universe, when the IGM is mostly neutral. Although simplistic,\footnote{This treatment neglects other cooling and heating processes that are relevant in the early universe, in particular, molecular cooling and interactions with CMB photons. These processes may be important in the formation of the first stars and galaxies; see e.g., \citet{Benson:2010aa}.} $T_{\mathrm{vir,c}}\sim10^{4}\,\mathrm{K}$ serves well enough to separate halos where gas can cool from those where it cannot, in the absence of molecular hydrogen or a photoionizing background.}

\apc{Following cosmic reionization, photoheating by the UV background increases the IGM temperature, making it it harder for low-mass halos to accumulate baryons. The UV background also acts to suppress radiative cooling in gas confined by halos. The combination of these two effects can be modeled as a rapid increase in the characteristic $T_{\mathrm{vir,c}}$ for galaxy formation after reionization \citep{Ikeuchi:1986aa, Rees:1986aa, Couchman:1986aa, Kauffmann:1993aa,Thoul:1996aa,Gnedin:2000aa, Benson:2002paper1, Benson:2002aa,Hoeft:2006aa,OkamotoGaoTheuns}. In detail, the effects of reionization on the accumulation and condensation of baryons are complex. The strength of the UV background and its interaction with the IGM are redshift dependent, and also density dependent; both galaxy formation and reionization proceed more rapidly in regions of higher density \citep[e.g.][]{Font:2011aa}. Predictions for this dependence are entangled with those for the rate of formation of the galaxies and quasars that give rise to the UV background. A self-consistent treatment requires radiative transfer and the resolution of sources of UV emission (and the surrounding interstellar medium) in low-mass halos, both of which are extremely computationally expensive in cosmological volume simulations.}

\apc{Hydrodynamical models of galaxy formation typically approximate reionization as an instantaneous and universal heating of the IGM. In semianalytic models, the effect of reionization on the confinement and condensation of gas can be parameterized as a threshold in halo virial velocity, $V_{\mathrm{cut}}$ (equivalent to $T_{\mathrm{vir,c}}$), below which no cooling takes place after $z_\mathrm{reion}$ \citep[e.g.][]{Benson:2003aa, Bower:2006aa}. This is the framework we use in this paper. Both $V_{\mathrm{cut}}$ and $z_\mathrm{reion}$ are usually taken to be universal parameters, motivated by the results of hydrodynamical simulations. \citet[][]{Font:2011aa} present a self-consistent semianalytic treatment of the evolution of the UV background, demonstrating that the effect of local reionization (for Milky Way--like dark matter halos) can be well approximated by adjusting $V_{\mathrm{cut}}$ and $z_\mathrm{reion}$.}

The effects described above imply the existence of a distinct population of `fossil' dwarf galaxies, associated with halos that exceed the cooling threshold before reionization but not afterward \citep{BKW, Benson:2002aa,Bovill_Ricotti2011,Font:2011aa,Bose:2018aa}. These fossil galaxies have been identified with the ultrafaint satellites of the Milky Way \citep[see, e.g.][]{McConnachie_local_group_2021,UDF_Simon_2019}. Simulations including realistic treatments of reionization predict bimodal satellite luminosity functions for Milky Way analogs, with one peak (very low luminosities but large numbers) corresponding to the fossils and a second peak (brighter but fewer in number) corresponding to halos that exceed $T_{\mathrm{vir,c}}$ after reionization \citep{Font:2011aa}. A fraction of dwarf galaxy host halos with particularly low late-time growth rates may pass below the cooling threshold for the first time at low redshift, suppressing their recent star formation \citep[e.g.][]{Pereira-Wilson:2023aa}. The properties and abundance of these populations, particularly the ultrafaint Milky Way satellites, currently provide the strongest observation constraint on the effective value of $V_{\mathrm{cut}}$, as well as a somewhat indirect constraint on $z_\mathrm{reion}$, complementary to measurements of the temperature and ionization of the IGM from the CMB and quasar absorption spectra \citep[e.g.][]{Bose:2018aa}. Predictions for the fossil/ultrafaint population have received considerable attention in the literature because they are also strongly affected by plausible variations in the dark matter power spectrum on scales that are otherwise unconstrained \citep[e.g.][and references therein]{Sawala_2015}.

\subsection{The ghostly galaxy scenario}

\citet{Benitez_Llambay_2020} describe the population of halos massive enough to retain their complement of baryons after reionization, but not massive enough to cool those baryons to the densities required for star formation at any epoch. Such halos are star-free but potentially gas-rich. Using a similar approach, we consider a different subset of halo merger trees, in which the \textit{main branch} (defined by the chain of most massive halo progenitors traced back from the halo at $z=0$) remains below $V_\mathrm{cut}$ at all epochs and hence does not host any in situ star formation, but  one or more \textit{minor branches} do exceed either the hydrogen cooling threshold (before reionization) or $V_\mathrm{cut}$ (after reionization), and hence can form stars. By construction, the stars formed in the minor branch are later accreted onto the star-free main branch.

\begin{figure}
    \includegraphics[width=\columnwidth]{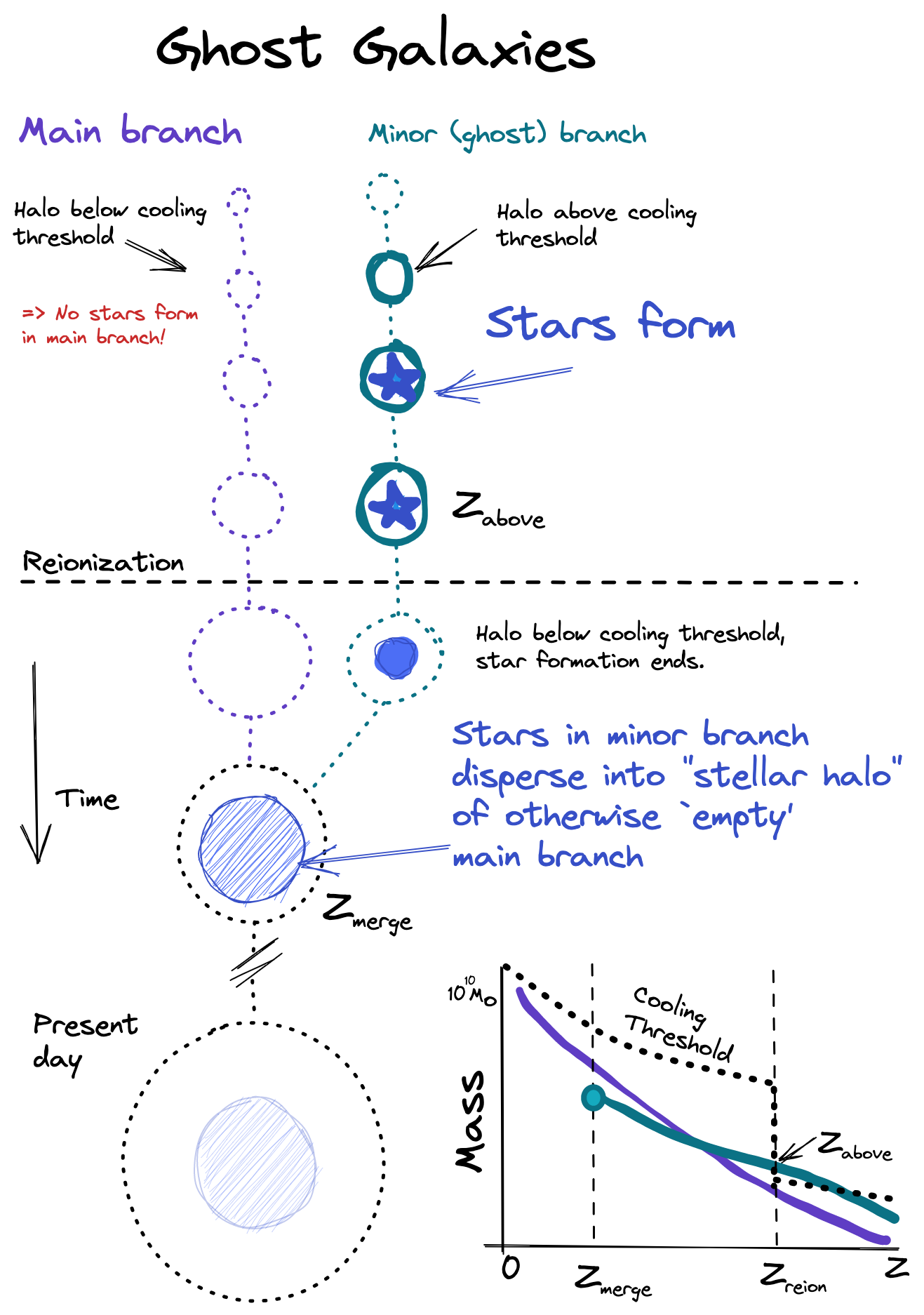}
    \caption{A cartoon of the `ghost galaxy' formation scenario. Time runs down the page. Dark matter halos above and below the hydrogen cooling threshold are indicated by solid and dotted circles, respectively, with radii indicating their mass. Two branches in the merger tree of a present-day halo merge at $z_\mathrm{merge}$. At this time, the less massive (minor) progenitor contains stars, whereas the more massive (main) progenitor does not. As sketched in the inset graph, this is possible if the minor branch grows faster at higher redshift, briefly exceeding the hydrogen cooling threshold up to a time $z_\mathrm{above}$ (thick lines). The main branch is always below this threshold. The (possible) result at the present day is a dark matter halo that is underluminous for its mass and contains only an LSB, stellar-halo-like component, comprising the stellar debris of the accreted minor branch.}
    \label{fig:ghost_cartoon}
\end{figure}

Fig.~\ref{fig:ghost_cartoon} is a cartoon of this scenario. The essential point concerns the fate of the stars formed in the minor branches, after they merge with the main branch. If a galaxy forms in the main branch, stars accreted from minor branches would constitute its `accreted stellar halo' at $z=0$. If a galaxy does not form \textit{in situ} in the main branch, those accreted `halo' stars will comprise the entirety of the stellar mass of the galaxy. Without a central concentration of in situ stars, those halos would not be distinguished as a separate structural component. Such objects would, in principle, have the characteristic features associated with stellar halos: high velocity dispersion, low concentration, and LSB \citep{Amorisco:2017aa_stellar_halos}. As we will confirm below, ghostly galaxies are plausible in $\Lambda$CDM, but expectations for their cosmic abundance, masses, and sizes are not at all obvious. Expectations for these properties will, almost by definition, be closely related to those for the stellar halos of regular dwarf galaxies, which have been studied in recent work by \citet{Kado-Fong:2020wg}, \citet{StellarHalo_Deason_2021}, and \citet[][, who also use the term `ghostly' to refer to such halos]{Ricotti:2022aa}. We consider this relationship further in section~\ref{sec:profiles}.

\section{Frequency of ghostly galaxies}
\label{sec:method}

To make quantitative predictions for the cosmic abundance of ghost galaxies, we apply two criteria (described below) to large numbers of halo merger trees. We obtain samples of merger trees using two different methods: a Monte Carlo approach, based on the extended Press--Schechter (EPS) formalism \citep{Lacey_Cole_1993} as implemented in the code of \citet{Parkinson2007},\footnote{\url{https://astro.dur.ac.uk/~cole/merger_trees/}} and a high-resolution $N$-body simulation, \citep[\coco{},][]{Hellwing_2016}, with a dark matter particle mass of $1.135\times10^{5}\,h^{-1}\,\mathrm{M_{\odot}}$ and Plummer-equivalent gravitational softening scale $230\,h^{-1}\,\mathrm{pc}$.

$N$-body simulations explicitly model the gravitational dynamics of structure formation and therefore provide more accurate (and detailed) predictions for halo mass assembly histories. Among other factors, they account for environmental effects \citep[e.g., on halo growth rates and structure; see][]{hellwing_coco_environment} and the survival of self-bound substructures within dark matter halos. However, the computational efficiency of the Press--Schechter approach allows for a much larger sample of trees to be constructed. This is particularly relevant in our case, because we are concerned with the smallest star-forming halos, which require a high-resolution (and hence necessarily small-volume) $N$-body simulation. Since we are studying a rare subset of these halos, restricting our analysis to the simulation alone would be a significant statistical limitation.

The EPS code provided by \citet{Parkinson2007} uses a Monte Carlo algorithm to generate merger trees consistent with a given initial matter power spectrum and cosmological parameters. We use the tabulated power spectrum from which the \coco{} initial conditions were generated and the same cosmological parameters as \coco{}. The \citet{Parkinson2007} algorithm includes additional tuning parameters to better match the statistics of EPS trees to those of trees obtained from $N$-body simulations. We set the values of these parameters following  Table 2 of \citet{Benson_2019}: $G=0.635, \gamma_{1}=0.176$ and $\gamma_{2}=0.041$.

The merger trees from the \coco{} simulation were constructed using the group-finding and `DHalo' linking procedures described by \citet{jiang2014_dhalos}. In detail, the algorithms used to identify bound structures in an $N$-body simulation and to link them between snapshots, are not straightforward. They require work-arounds for the effects of limited numerical resolution, which may differ between group-finding algorithms (e.g., the difficulty of identifying subhalos as they pass through the centers of their hosts can create artificial `breaks' in trees). They may also involve somewhat arbitrary choices (e.g., regarding the treatment of subhalos that escape their hosts). The DHalo procedure is designed to be robust against many of these issues. Nevertheless, there may be edge cases that have not been accounted for, which may be more apparent toward the resolution limit and in higher-resolution simulations such as \coco{}. This provides more motivation for our comparison with Press--Schechter trees, which, although more limited in some respects, have the advantage of a clear and consistent operational definition.  

\subsection{Identifying ghosts}
\label{sec:define_ghosts}

A merger tree comprises a set of \textit{nodes} (representing virialized dark matter halos) identified at a series of discrete timesteps (\textit{snapshots}) ranging from $z=0$ to $z\sim20$. Nodes are linked by pointers to their \textit{descendant} (one-to-one, forward in time) and \textit{progenitors} (one-to-many, backward in time). We identify trees associated with ghost galaxies by traversing these pointers and applying the two criteria described below to all the nodes in the tree.

\subsubsection{Main branch criterion}

Each tree has a single \textit{root node}, which corresponds to an isolated dark matter halo at $z=0$. The main branch of a tree is defined by the chain of most massive progenitors traced backward in time from the root node. Here `mass' refers to the total virial mass of the system (baryons and dark matter), taken to be equivalent to $M_{200}$, the mass enclosed by a density contour at 200 times the critical density for closure. The main branch, in principle, corresponds to the central potential that dominates the system at $z=0$.

\updated{A necessary condition for the formation of a ghost galaxy is that no stars should form in a cooling flow that would produce a compact stellar system deeply embedded in the present-day potential (i.e. there should be no `in situ' component). Our \textit{first criterion} is therefore that no main branch node should exceed the following virial temperature thresholds:
\begin{align}
    & T_{200,\mathrm{cut}} \sim 10,000\,\mathrm{K} & (z>z_{\mathrm{reion}}) \label{eq:tcrit_before_reion} \\ 
    & T_{200,\mathrm{cut}} \sim 32,000\,\mathrm{K} & (z<z_{\mathrm{reion}}) \label{eq:tcrit_after_reion} 
\end{align}
where throughout we take $z_{\mathrm{reion}}=10$ as fiducial choice of the redshift of reionization.  As described above and in more detail by \citet{Benitez_Llambay_2020}, these thresholds correspond to the limits on cooling imposed (before reionization) by the temperature at which atomic hydrogen in equilibrium with dark matter halos can be ionized and (after reionization) by the higher temperature of the IGM due to the cosmic UV background, which reduces the cooling efficiency of virialized gas and prevents baryons from accreting onto low-mass halos. The exact values of the thresholds depend on a number of assumptions about the thermal physics of the IGM and the virialized gas and about the strength and effects of the ionizing background.}

\updated{We implement these thresholds in the same way as the Galform semianalytic model \citep{Cole_Lacey_2000_Galform,Bower:2006aa, Font:2011aa, Lacey:2016aa}, parameterized by $z_{\mathrm{reion}}$ and a threshold circular velocity, $V_{\mathrm{cut}}$, which encapsulates the effects of IGM heating \citep{OkamotoGaoTheuns}. At $z<z_{\mathrm{reion}}$, no cooling, and hence no star formation, can occur in halos with virial velocity $V_{200} = (GM_{200}/R_{200})^{1/2} < V_{\mathrm{cut}}$. Following \citet{Font:2011aa}, we assume a fiducial value of $V_\mathrm{cut} = 30\,\mathrm{km\,s^{-1}}$, which corresponds to $T_{200,\mathrm{cut}}\simeq31,650\,\mathrm{K}$. At $z>z_{\mathrm{reion}}$, we assume a fiducial cooling floor of exactly $10 000\,\mathrm{K}$. In later sections, we explore variations of these values.}

\subsubsection{Minor branch criterion}
\label{sec:minor_branch_and_defns}

\begin{figure}
    \includegraphics[width=1.1\columnwidth]{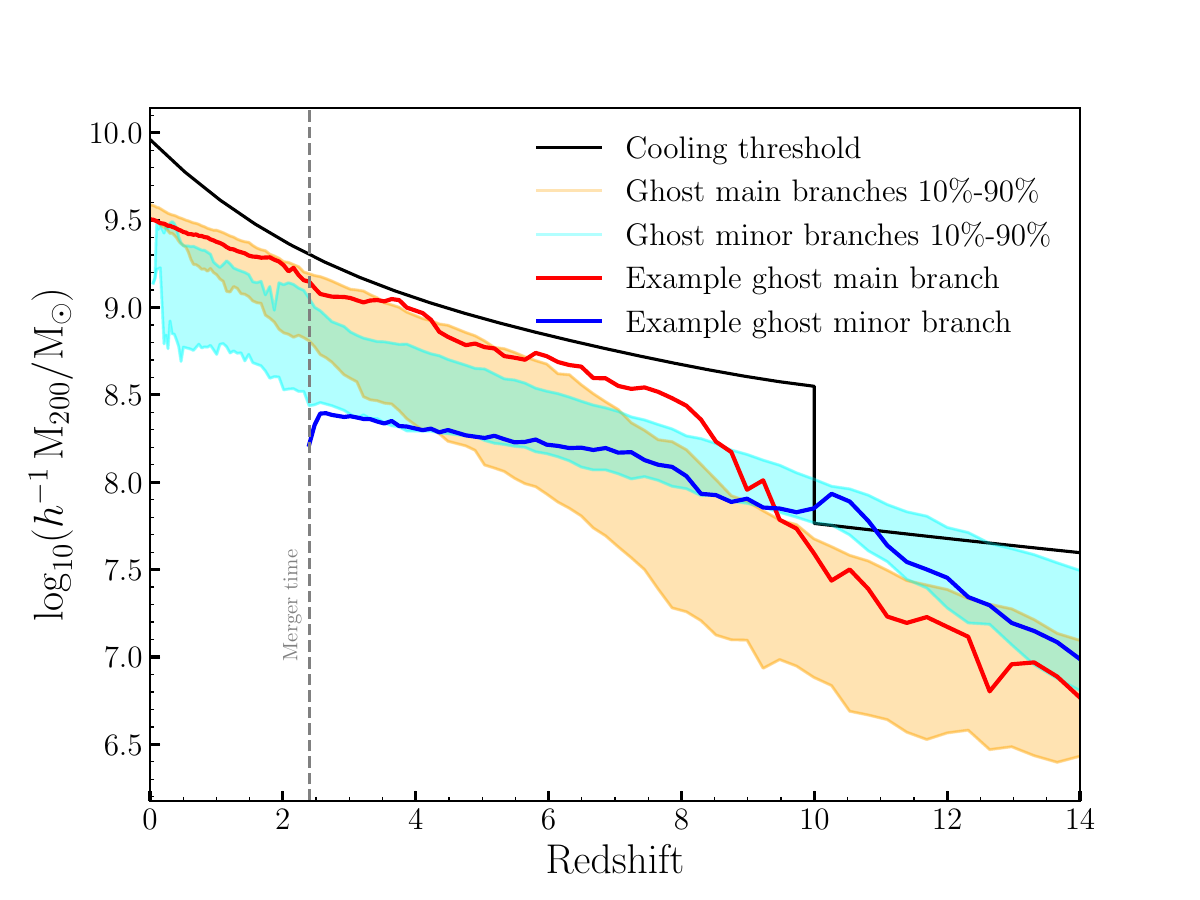}
    \caption{Mass growth histories of relevant merger tree branches from the \coco{} simulation with $M(z=0)\simeq10^{9.5}\,\mathrm{M_{\odot}}$.  Red and blue solid lines show the evolution of the main and minor branches, respectively, in a randomly chosen tree of this mass that meets our `ghost galaxy' criteria. The solid black line shows the halo mass corresponding to the cooling threshold described in the text. The sharp increase in the threshold mass at $z=10$ corresponds to our fiducial model of reionization ($V_\mathrm{cut}=30\,\kms$). We associated this tree with a ghost galaxy at $z=0$, because the minor branch crosses the cooling threshold (up to $z_\mathrm{above}=10$, $m_\mathrm{above}\sim10^{8}\,h^{-1}\mathrm{M_{\odot}}$) while the main branch does not. The two branches merge at $z_\mathrm{merge}\approx2.4$ (dashed line), at which point their mass ratio is $m_{\mathrm{merge}}/M_{\mathrm{merge}}\sim10\%$. Shaded areas show the 10$^{\mathrm{th}}$--$90^{\mathrm{th}}$ percentile range of mass histories for main branches (red) and minor branches (blue) in all such trees, for this choice of final mass.}
    \label{fig:finalHM_hist_figure}
\end{figure}

\begin{figure}
    \centering
    \includegraphics[width=\columnwidth,trim=2.5cm 0cm 3cm 0, clip=True]{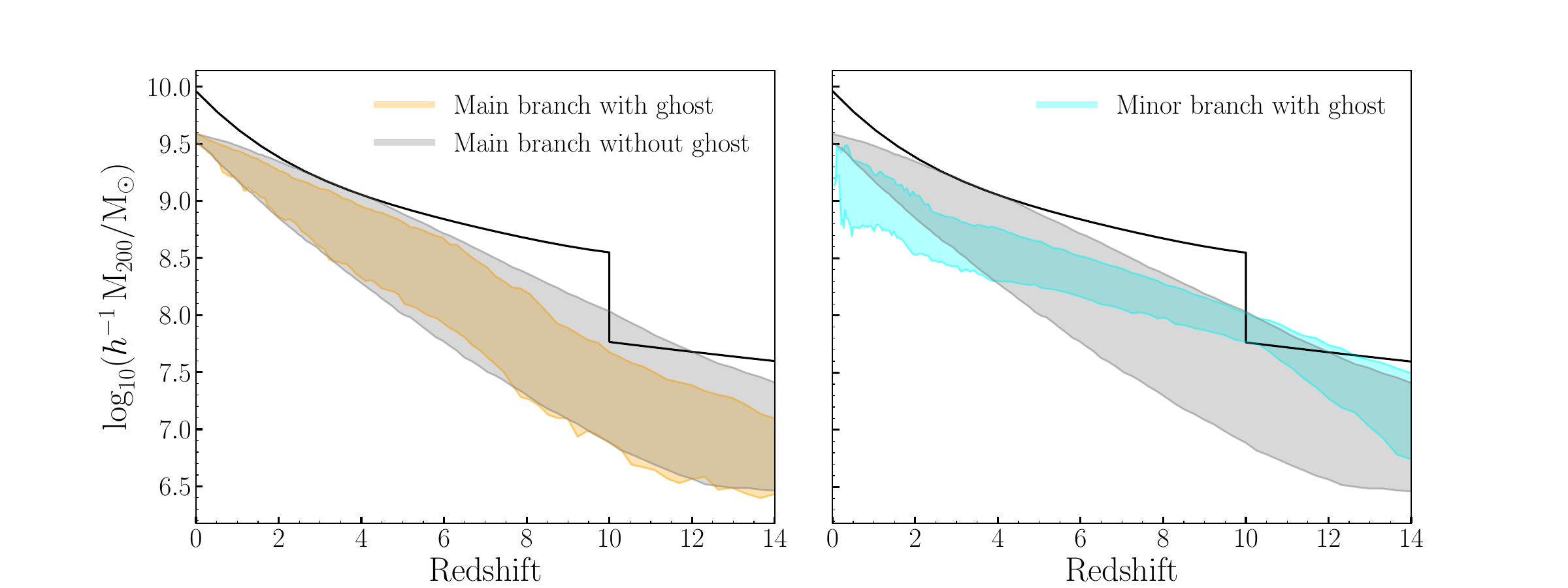}
    \caption{The shaded gray region (same in both panels) shows the 10$^{\mathrm{th}}$ -- 90$^{\mathrm{th}}$ percentile range of mass growth histories for the main branches of all halos in the \coco{} simulation with $M(z=0)\simeq10^{9.5}\,\mathrm{M_{\odot}}$.
    Shaded orange and blue regions, repeated from Fig.~\ref{fig:finalHM_hist_figure}, show the corresponding range for main (left) and minor (right) branches in the subset of these halos that host ghost galaxies.}
    \label{fig:ShadedArea.pdf}
\end{figure}

Where a main branch node has more than one progenitor, the less massive progenitors correspond to the end points of minor branches. Each minor branch merging onto the main branch comprises an independent hierarchy of less massive progenitor branches. In the Press--Schechter formalism, minor branches are interpreted as losing their separate identity as soon as they merge into a more massive branch.

Our \textit{second criterion} for a tree to be associated with a ghost galaxy is therefore that star formation can occur in \textit{at least one} of its minor branches, i.e. that at least one node in any minor branch exceeds the cooling threshold (Equation~\ref{eq:tcrit_before_reion} or \ref{eq:tcrit_after_reion}, depending on the redshift of the node). 

We do not require that this occur before reionization, that it occur in only one minor branch (there may be multiple ghost galaxy progenitors in a single tree), or that it occur only in minor branches merging directly onto the main branch, rather than at deeper levels of the hierarchy. 

In this simple formulation, all the stellar mass that comprises the ghost galaxy at $z=0$ forms in the minor branch when they exceed the cooling threshold. Later, when those branches merge onto the main branch of their tree, those stars are distributed on weakly bound orbits in the main branch potential. 

Note that it is not a contradiction for a minor branch node to be more massive than the main branch node at any snapshot, except the snapshot immediately before the two branches merge. In such cases, the main branch progenitor (by construction) must subsequently grow more rapidly, such that it is more massive when the two branches merge.

Our two criteria introduce the following characteristic times (redshifts) and masses, which we refer to throughout the paper:

\begin{itemize}
    \item $z_\mathrm{above}$, the \textit{lowest} redshift (latest time) at which a minor branch exceeds the cooling threshold;
    \item $m_\mathrm{above}$, the mass of the minor branch at $z_\mathrm{above}$;
    \item $z_\mathrm{merge}$, the redshift at which the minor branch is last identified in the merger tree;
    \item $m_\mathrm{merge}$, the mass of the minor branch at $z_\mathrm{merge}$;
    \item $M_\mathrm{merge}$, the mass of the main branch at $z_\mathrm{merge}$.
\end{itemize}

In practice, the mass associated with minor branches may survive as a self-bound subhalo orbiting within the virial radius of the main branch for some time after $z_\mathrm{merge}$. This is ignored in the Press--Schechter formalism but modeled explicitly in the $N$-body case. The definition of `DHalos' in the tree-building procedure of \citet{jiang2014_dhalos} attempts to match the Press-Schechter definition of a ``halo.'' We leave this complication aside for now and consider only `halos' in the Press--Schechter sense. Later we explore merger trees based on subhalos, rather than DHalos. Operationally, halo mass is defined within an overdensity contour 200 times the critical density for closure, which we treat as a proxy for virial mass.

\subsubsection{Main and minor branch growth histories}

Fig.~\ref{fig:finalHM_hist_figure} shows an example of the growth history of the main branch and the minor branch in the $N$-body merger tree of a halo that we identify with a potential ghostly galaxy. The solid black line shows the virial mass equivalent to $T_\mathrm{crit}$, the threshold virial temperature for cooling, which reaches $M_{\mathrm{crit}} \approx 10^{10}\,h^{-1}\mathrm{M_{\odot}}$ at $z=0$. The sharp transition in this curve at $z=10$ corresponds to our fiducial treatment of reionization, as described above. The main branch of the example tree (red line) corresponds to a system with present-day virial mass $M_{200}(z=0)\simeq10^{9.5}\,h^{-1}\mathrm{M_{\odot}}$. Its mass\footnote{Mass growth histories in $N$-body simulations may not be strictly monotonic, as in this case. This is one of the many issues with the implementation of $N$-body tree-building algorithms we refer to in Section 2.} is less than $M_{\mathrm{crit}}$, not only at $z=0$ but at all redshifts, in line with our first criterion. Conversely, one of the minor branches of this tree (blue line) grows more quickly than the main branch progenitor at high redshift and briefly exceeds the cooling threshold before reionization ($z_\mathrm{above}\simeq z_\mathrm{reion} = 10$). The two branches merge at $z_\mathrm{merge}\approx2.4$ (dashed line). The mass ratio of the two branches at $z_\mathrm{merge}$ is approximately 10:1. 

For comparison, the shaded regions in Fig.~\ref{fig:finalHM_hist_figure} show the envelope of histories for main branches and minor branches in all \coco{} $N$-body trees with $M_{200}(z=0)\simeq10^{9.5}\,h^{-1}\mathrm{M_{\odot}}$ that meet both our ghost galaxy criteria. The random example we have chosen exaggerates the difference in the growth rates of the two branches at $z<10$, but is otherwise typical. Note that the main branch distribution narrows toward $z=0$ by construction, whereas the minor branch distribution broadens as the number of surviving minor branches decreases. It is clear that, for this choice of final mass, effectively all the minor branches exceed the threshold mass at $z>10$, and not at lower redshift. Note also, however, that $M_{200}(z=0)$ in this example is lower than the maximum of $\simeq 10^{10}\,M_{\odot}$ set by our main branch criterion. 

The left panel of \updated{Fig.~\ref{fig:ShadedArea.pdf}  contrasts the formation histories of ghost galaxy main branches (orange) and those of other halos with the same present-day virial mass, in this case $M_{200}(z=0)\simeq10^{9.5}\,\mathrm{M_{\odot}}$. The gray envelope includes main branch histories that cross the threshold, before reionization and at later times up to $z\sim2$ (these are a very small population for this choice of present-day mass), as well as histories that never cross the threshold. The orange envelope of the ghost main branches shows clearly that they are drawn from the latter population. The blue region in the right panel shows the corresponding growth rates of the star-forming minor branches of the ghosts. Essentially by construction, these minor branches are more massive than the ghost main branches at high redshift but have substantially slower growth rates and hence lower masses at low redshift\footnote{\apc{We note that these differences in assembly history imply that, at higher present-day halo masses, the halos of ghosts may be increasingly extreme outliers in the concentration-mass relation.}}.}

\subsection{Ghost galaxy fractions}

\begin{figure}
	\includegraphics[width=\columnwidth]{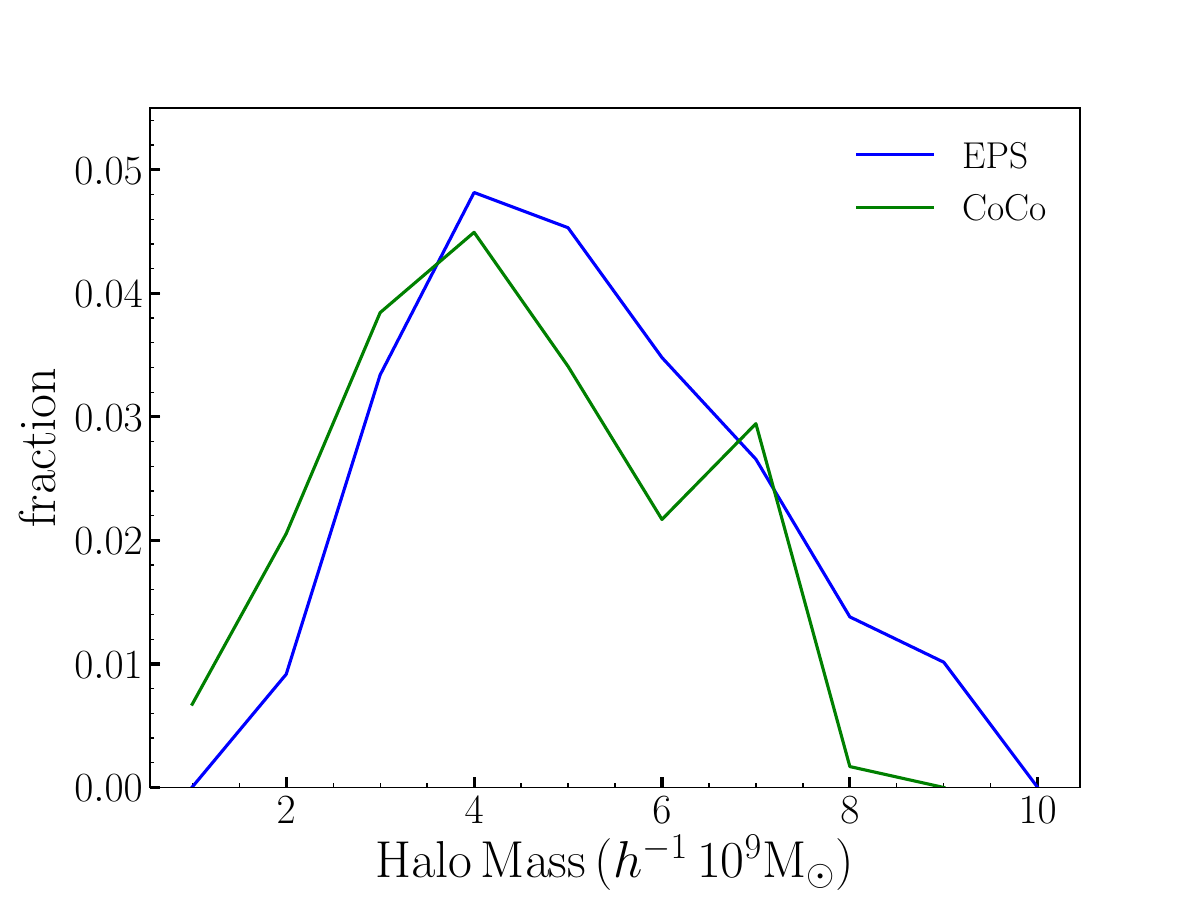}
    \caption{The fraction of merger trees for dark matter halos of a given mass that host ghost galaxies at $z=0$, for $\mathrm{V_{\mathrm{cut}}}=30\,\kms$. The result based on EPS trees is shown in green, and the result based on trees from the \coco{} simulation is shown in blue. The probability of finding ghosts in the field peaks at $\sim5\%$ around a halo mass $\approx4\times10^{9}h^{-1}\,\mathrm{M_{\odot}}$.}
    \label{fig:abovefrac_cocoEPS}
\end{figure}

\begin{figure}
    \includegraphics[width=\columnwidth]{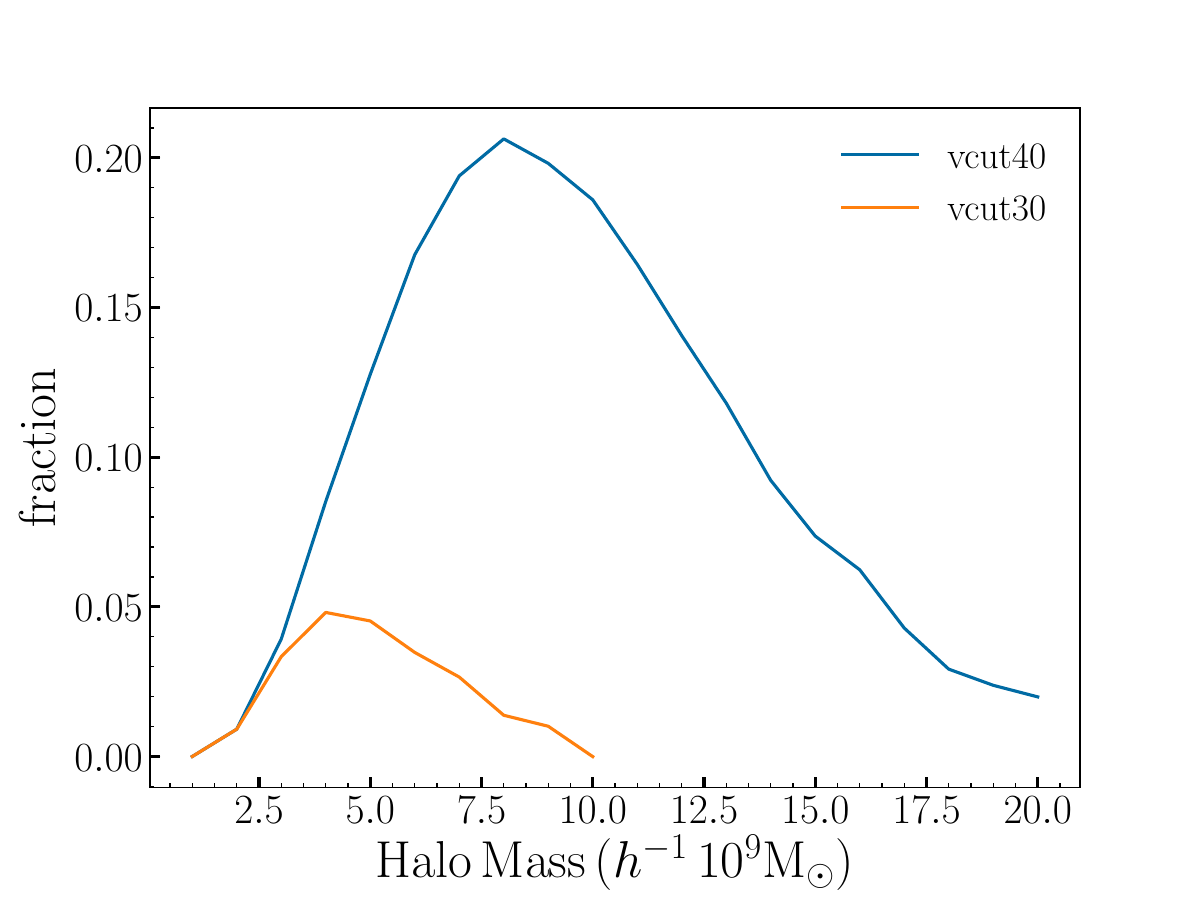}
    \caption{The effects of changing $\mathrm{V_{\mathrm{cut}}}$ from $30\,\kms$ (orange line, repeating Fig.~\ref{fig:abovefrac_cocoEPS}) to  $40\,\kms$ (blue line). A higher cooling threshold increases the peak fraction of ghost galaxies (to, e.g., $\sim20\%$ of all halos at $\approx8\times10^{9}h^{-1}\,\mathrm{M_{\odot}}$) and broadens the mass range of hosts.}
    \label{fig:abovefrac_vcut3040}
\end{figure}

Fig.~\ref{fig:abovefrac_cocoEPS} shows how often the above criteria are satisfied as function of the virial mass of the main branch at $z=0$. We determine the fraction $f_\mathrm{ghost} = N_\mathrm{ghosts}/N_\mathrm{total}$ in mass bins of width $10^{9}\,h^{-1}\,\mathrm{M_{\odot}}$. For the EPS trees, this fraction is computed using $N_\mathrm{total}= 100,000$ trees of the same final mass. For \coco{}, final masses are distributed across the mass bins according to the halo mass function in the simulation volume (with 17,642 trees in the range $1\times 10^{9}< M_{200} < 2\times10^{9} \,h^{-1} \,\mathrm{M_{\odot}}$ and 482 trees in the range $9\times10^{9}< M_{200} < 1\times 10^{10}\,h^{-1} \,\mathrm{M_{\odot}}$). Cases in which the ghost galaxy at $z=0$ has multiple minor branch progenitors are counted several times in Fig.~\ref{fig:abovefrac_cocoEPS}. We discuss the treatment of multiple ghost progenitors further below in the context of predictions for the stellar mass function.

Fig.~\ref{fig:abovefrac_cocoEPS} shows that, with our fiducial treatment of reionization, ghosts are most likely in halos of present-day mass $4-5\times10^{9}\,h^{-1}M_{\odot}$, in which range $f_\mathrm{ghost} \simeq 5\%$. The peak in the distribution reflects the interaction of our two criteria. The fraction of main branches meeting the first criterion falls with increasing final mass, whereas the fraction of minor branches meeting the second criterion rises with increasing final mass.  The results from the much smaller sample of trees in \coco{} agree well with the EPS predictions, perhaps showing a small offset toward lower mass. We conclude that EPS trees provide a sufficiently robust description of the $N$-body results for our purposes in this paper;\footnote{\updated{The EPS approach requires tuning to reproduce the mass assembly histories of $N$-body trees accurately. We examined an alternative version of our EPS code with additional parameters to improve the overall match to the conditional mass functions of low-mass progenitors in \coco{} at high redshift. However, although this change improves the overall correspondence between the EPS and $N$-body trees, we found that it significantly underpredicts the (already small) number of progenitors with the highest mass ratios at high redshift in \coco{}. This has a particularly strong impact on our predictions for ghost galaxies in \coco{}.}} since the EPS method provides much larger samples, we refer mainly to the EPS results in the following discussion. We will discuss dynamical insights from the $N$-body trees in Sections~\ref{sec:merger_mass_ratios_and_times} and \ref{sec:profiles}.

\subsubsection{Changing $V_\mathrm{cut}$}

We now consider how variations in the parameters in our simplified model of the cooling threshold ($V_\mathrm{cut}$) and its redshift evolution ($z_\mathrm{cut}$) affect predictions for the fraction of ghost halos at a given $z=0$ virial mass.

Fig.~\ref{fig:abovefrac_vcut3040} shows equivalent results for an alternative model of reionization with $V_\mathrm{cut}=40\,\kms$, in which baryon accretion and cooling are suppressed in more massive halos after reionization. A higher $V_\mathrm{cut}$ may correspond either to a more intense cosmic UV background overall or to the more local enhancement of the UV background in dense regions (in which case the effective value of $z_\mathrm{reion}$ would also be higher, but we ignore that here for simplicity). \citet{Font:2011aa} found $V_\mathrm{cut}=30\,\kms$ to be an appropriate value for the simplified global reionization model we use here. They determined this by comparing the observed satellite luminosity function of the Milky Way to the predictions of a more detailed semianalytic model of heating by the local and global cosmic UV background. This calibration may be sensitive to other aspects of the treatment of galaxy formation in the model, such as the strength of feedback and the escape fraction of ionizing photos, and also to uncertainties in the Milky Way's satellite luminosity function and total mass.

There is a striking difference between our predictions for $V_\mathrm{cut}=30\,\kms$ (repeated for reference in Fig.~\ref{fig:abovefrac_vcut3040} as an orange line) and those for $V_\mathrm{cut}=40\,\kms$. The fraction of trees resulting in ghost galaxies for $V_\mathrm{cut}=40\,\kms$ peaks at $f_\mathrm{ghost} \simeq 20\%$, around $\simeq9\times10^{8}\,h^{-1}M_{\odot}$, and has a much broader distribution, with a tail to $\sim3\times10^{10}\,h^{-1}\mathrm{M_{\odot}}$. 

Stronger IGM heating (i.e. a higher threshold mass at $z<z_\mathrm{reion}$) makes it less likely that main branches of a given final mass will meet our first criterion,\footnote{See section 4 of \citet{Benitez_Llambay_2020} for discussion of the effects of $z_\mathrm{reion}$ and $V_\mathrm{cut}$ on the fraction of main branches hosting star formation.} but it does not affect the probability of minor branches meeting our second criterion at $z>z_\mathrm{reion}$. Of course, stronger heating greatly reduces the fraction of minor branches that exceed the threshold at $z<z_\mathrm{reion}$. Consequently, as we discuss in the next section, stronger reionization creates more ghosts and associates them with more massive halos (and hence, potentially, more extreme surface brightnesses), but it also limits their maximum luminosity. 

\begin{figure}
	\includegraphics[width=\columnwidth]{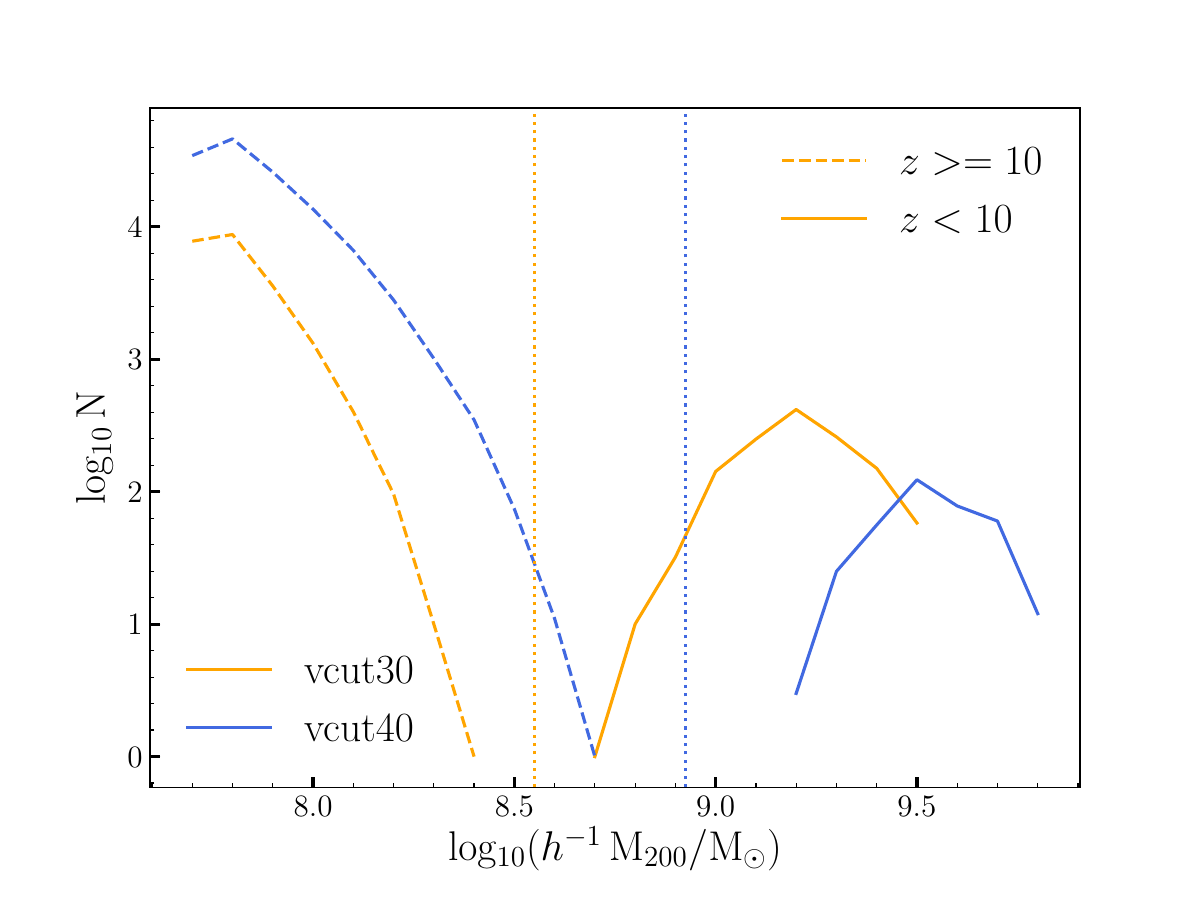}
    \caption{Histogram of minor branch halo masses $m_\mathrm{above}$, measured at $z_\mathrm{above}$, the lowest redshift (latest time) at which the branch exceeded the cooling threshold mass \reply{for $V_\mathrm{cut}=30$ and $40\,\mathrm{km\,s^{-1}}$, in orange and blue, respectively. The dotted lines show the corresponding threshold masses for star formation at $z_\mathrm{reion}=10$. Dashed and solid lines correspond to minor branches that cross the threshold before and after $z_\mathrm{reion}$, respectively}. The bimodality arises because there is a population of halos that exceeds the threshold before reionization at $z_\mathrm{reion}$ but not afterward. The sharp lower mass limit corresponds to the least massive halos that exceed the threshold at $z_\mathrm{reion}$.  The gap is created by the large instantaneous increase in the threshold at $z_\mathrm{reion}$ (see  Fig.~\ref{fig:finalHM_hist_figure}) as explained in the text.}
    \label{fig:last_aboveHM_vcut30} 
\end{figure}

Fig.~\ref{fig:last_aboveHM_vcut30} shows (for our fiducial $V_\mathrm{cut}=30\,\kms$ model) the histogram of $m_\mathrm{above}$, the halo masses of the star-forming minor branches in the ghost galaxy trees at $z_\mathrm{above}$ (when they were last above the cooling threshold mass). In the next section, we will use a simple function of this mass to assign stellar masses to the ghosts. We see two peaks in this halo mass distribution, corresponding to branches in which star formation is truncated by reionization (blue) and branches that exceed the cooling threshold at lower redshift (orange). The gap corresponds to halos that find themselves crossing below the instantaneously increased threshold at $z_\mathrm{reion}$, but grow above it again at a lower redshift ($z\sim7$ for the most massive). For these branches, as the mass of the branch at $z_\mathrm{reion}$ approaches the threshold mass at that time, it becomes increasingly unlikely that the branch will not cross above the threshold again later and reach a much greater maximum mass. The high-mass peak is modulated, however, by the requirement of merging with a permanently dark main branch\footnote{In the case of Milky Way satellites, discussed in the next paragraph, the modulating effect on the high-mass part of the distribution is the requirement of merging with a Milky Way--mass halo.}. 

A similar bimodality arises in  predictions for the Milky Way satellite luminosity function \citep[e.g.][]{Bovill:2009aa,Li:2010aa,Font:2011aa,Bose:2018aa}. In that context, galaxies associated with halos that do not accrete or cool gas after reionization are usually called `ultra faints' or `reionization fossils' \citep{Bovill:2009aa}. Most known ultrafaints are satellites of the Milky Way and M31, although it is likely that they also exist in the field \citep[e.g.][]{sand:2022aa_tucb}. Although essentially all ghost galaxies are fossils, not all fossil galaxies are ghosts. Stars in typical fossil galaxies should be deeply embedded in a potential similar to that in which they formed and hence should have a compact density profile; conversely, stars in ghosts are expected to comprise dynamically hotter, more diffuse systems at $z=0$, because they form in branches that merge into the `dark' central potential \citep[e.g.][]{Amorisco:2017aa_stellar_halos}. 

The higher-mass peak is absent for $V_\mathrm{cut}=40\,\kms$, and the amplitude of the lower-mass peak increases, for reasons discussed above. In a model with weaker IGM heating (e.g.\  $V_\mathrm{cut}=20\,\kms$), minor branches can cross the threshold more easily after reionization, but the present-day mass range of trees satisfying the first criterion (main branch never cools) greatly reduces the number of ghost trees overall, as well as the maximum masses of minor branches associated with those trees. The overall qualitative result is that ghost galaxies are not expected in significant numbers for $V_\mathrm{cut}<30\,\kms$ and any that do form are unlikely to be detectable. The absence (or low abundance) of ghost galaxies would therefore imply weak IGM heating during reionization. Conversely, large numbers of faint ghost galaxies would imply stronger reionization, either globally or locally in particular regions. This simple picture is, of course, subject to a great deal of uncertainty regarding the luminosity and structure of ghost galaxies, their detectability, and the ease with which they can be separated from `ordinary' dwarf galaxies.

\updated{\citet{Benitez_Llambay_2020} use a hydrodynamical simulation to calibrate a more complete semianalytic model for the evolution of the characteristic temperature of the IGM accreted by halos after reionization. This model is equivalent to a redshift-dependent variation in $V_\mathrm{cut}$ from $\approx20\,\mathrm{km\,s^{-1}}$ at $z=10$ to $\approx25\,\mathrm{km\,s^{-1}}$ at $z=0$. We have  examined the abundance of ghosts with this redshift-dependent $V_\mathrm{cut}$, fixing $z_\mathrm{reion}=10$. The lower mean value of $V_\mathrm{cut}$ reduces the typical halo mass of the ghosts and the dispersion around that mass, as discussed above. The modest increase of $V_\mathrm{cut}$ with redshift allows relatively more minor branches in ghost trees to host star formation at lower redshift but makes it harder for main branches to remain below the threshold. The overall effect is to boost the fraction of ghosts among the least massive halos in the mass range. The fractions in this case are comparable to those in our fiducial $V_\mathrm{cut}=30\,\mathrm{km\,s^{-1}}$ model (up to $\sim10\%$ at $M_{200}\sim2\times10^{9}\,h\,\mathrm{M_{\sun}}$).}

\updated{The cooling threshold before reionization can also be varied. Our fiducial choice of $T_\mathrm{200,cut}=10,000\,\mathrm{K}$ is equivalent to $V_\mathrm{cut}\simeq17\,\mathrm{km\,s^{-1}}$. We do not explore changes in $T_\mathrm{200,cut}$ in detail because we consider its uncertainty to be less significant than that associated with the treatment of reionization. For example, \citet{Benitez_Llambay_2020} use $V_\mathrm{cut}\simeq13\,\mathrm{km\,s^{-1}}$ ($5700\,\mathrm{K}$) for their preferred model. In general, lower $T_\mathrm{200,cut}$ reduces the number of star-free main branches, while higher values increase the number of dark main branches but reduce the number of star-forming minor branches.}

\subsubsection{Changing $z_\mathrm{cut}$}

\updated{The effective global redshift of reionization (defined as the epoch at which the ionized fraction is 50~per~cent) is restricted by observations to an interval $8 \lesssim z_\mathrm{reion} \lesssim 10$ \citep[e.g.][]{planck_reionization_2016}. Our $z_\mathrm{cut}$ parameter corresponds to the time at which gas accretion and cooling are significantly affected by the UV background. These definitions are similar but not identical. The uncertainty in the choice of $z_\mathrm{cut}$ is therefore greater than the uncertainty in observational estimates of $z_\mathrm{reion}$.}

\begin{figure}
	\includegraphics[width=\columnwidth]{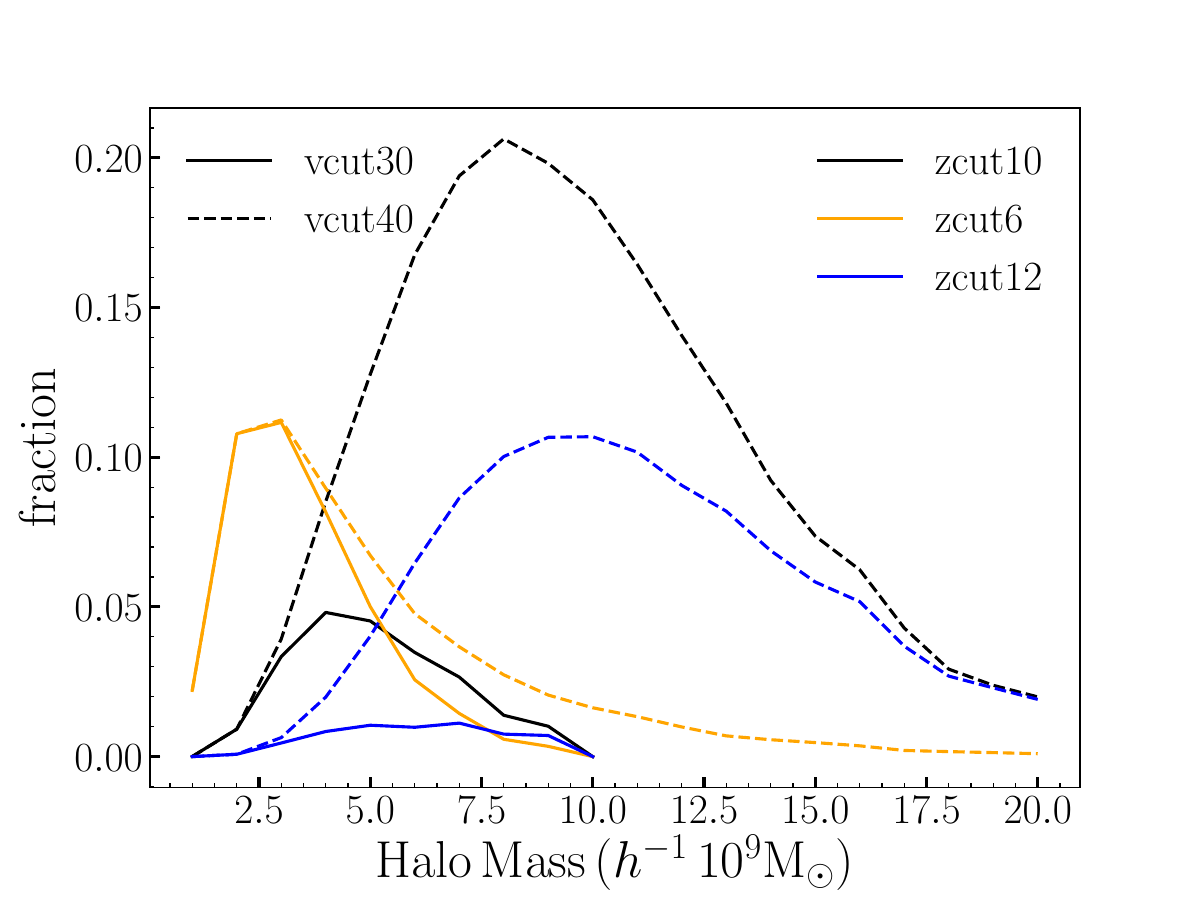}
    \caption{Same as Fig.~\ref{fig:abovefrac_vcut3040}, but showing the effects of varying $z_\mathrm{cut}$.}
    \label{fig:vary_zcut}
\end{figure}

\updated{Fig.~\ref{fig:vary_zcut} shows how different choices of $z_\mathrm{cut}$ change the results shown in Fig.~\ref{fig:abovefrac_vcut3040}. Relative to our fiducial choice of $z_\mathrm{cut} = 10$, assuming $z_\mathrm{cut} = 6$ greatly reduces the typical mass of halos hosting ghost galaxies at $z=0$, and makes the fraction of these halos at a given mass less sensitive to the value of $V_\mathrm{cut}$. Adopting $z_\mathrm{cut} = 12$ increases the typical host mass slightly but also reduces the fraction of hosts at a given mass. The reduction is greater for lower $V_\mathrm{cut}$.}

\updated{These effects follow from the two requirements that define ghost galaxy hosts and the typical shape of halo mass accretion histories. If $z_\mathrm{cut}$ is low, all but the least massive and slowest-growing main branches pass above the threshold at high redshift. The ghost population is maintained by the increase in the fraction of (low-mass) minor branches associated with these trees that can pass above the threshold. Conversely, if $z_\mathrm{cut}$ is high, more main branches remain below the threshold, but fewer low-mass minor branches exceed it. This depletes the (dominant) population of ghosts associated with minor branches that are truncated by reionization. The fraction of the most massive halos hosting ghosts is still determined by $V_\mathrm{cut}$ at relatively low redshift and hence less affected by a higher $z_\mathrm{cut}$. Our fiducial choice of $z_\mathrm{cut}=10$ therefore (approximately) maximizes the fraction of relatively massive halos that contain ghost galaxies. }
 
\subsection{Estimating the stellar mass function}

To investigate the cosmic abundance of ghostly galaxies and the potential for their detection in surveys, we use a simple prescription to derive their stellar mass function from their halo mass. For this analysis we  only consider trees generated with the EPS method, which yields predictions comparable to those of our $N$-body simulation but for a much larger sample of trees. In linearly spaced halo mass bins of width $1\times10^{9}$, we generate 1 million EPS trees. We assign each tree a fractional weight in order to recover the same volume density of halos as \coco{} in the same mass bin. The halo mass function of ghost galaxies is then simply the convolution of the halo mass function with the results shown in Fig.~\ref{fig:abovefrac_vcut3040}. 

The star formation efficiencies of low-mass halos at $z\gtrsim3$ are highly uncertain. Fits to observed luminosity functions based on variants of the abundance matching ansatz, as in \citet{Behroozi_2019}, suggest that the average stellar mass--halo mass (SMHM) relation is redshift dependent. High-resolution hydrodynamical simulations of low-mass halos \citep{Sawala_2015} also show considerable scatter in their star formation efficiencies. The scatter reflects the stochastic assembly and thresholds on cooling we have described above and is likely increased by the complex interaction of star formation and feedback. 

Given this uncertainty, we prefer to \reply{take a simple}, easily understood and reproducible approach that can be updated in more detailed work, or when better constraints on high-redshift star formation are available. We assume that the stellar mass associated with a halo can be derived from the virial mass $m_\mathrm{above}$ (defined at $z_\mathrm{above}$, the latest time at which the merger tree branch associated with the ghost exceeds the cooling threshold). We use two methods to convert $m_\mathrm{above}$ to a stellar mass, $M_\star$. The first is to assume that a universal fraction of available baryons is converted to stars, with no dependence on $m_\mathrm{above}$ or $z_\mathrm{above}$. The second is to obtain a star formation efficiency from the redshift-dependent SMHM relations given in Appendix J of \citet{Behroozi_2019}. These functions yield a formation efficiency that increases with redshift and with mass at a fixed redshift (in the low-mass regime).  We caution that the \citet{Behroozi_2019} SMHM relations are not well constrained below $M_{200}\sim10^{8}\,\mathrm{M_{\odot}}$ or at redshifts $3\lesssim z \lesssim 10$. They are essentially unconstrained at higher redshift. These are the regimes of interest for ghost galaxies. Recent work by  \citet{Wang2021} showed the linear extrapolation of the low-mass end of the SMHM relation from UniverseMachine agrees with the Milky Way satellites but the inferred star formation histories do not. Contrasting this poorly constrained but somewhat more `realistic' approach with the simplistic assumption of a fixed efficiency illustrates the basic effects of redshift and mass dependence. 

\begin{figure}
    \includegraphics[width=\columnwidth]{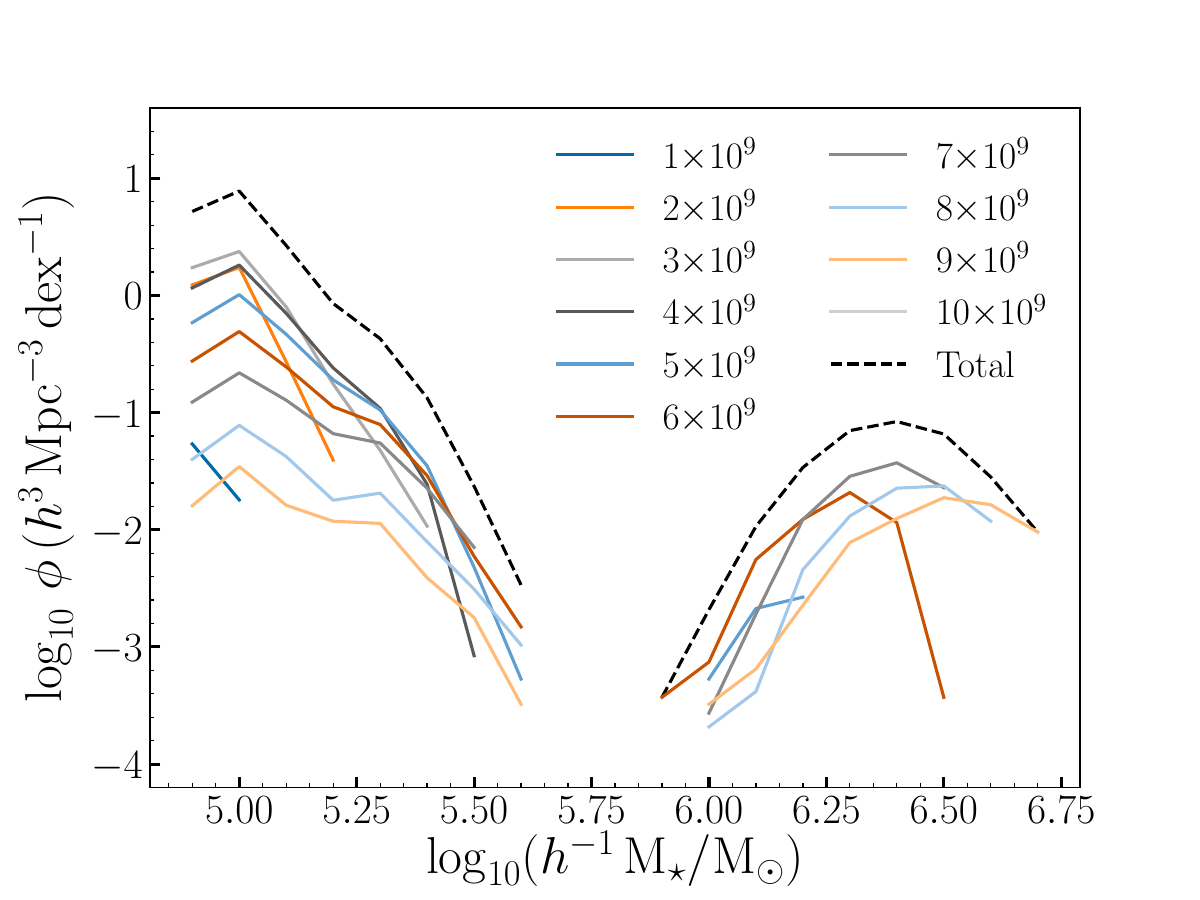}
    \caption{The stellar mass function of ghost galaxies for our $V_\mathrm{cut}=30\,\kms$ model, assuming 1\% of available baryons are converted to stars at $z_{\mathrm{above}}$. Colors show the contribution of different bins of present-day virial mass, in the range  $10^9 - 10^{10}\,\mathrm{M_{\odot}}$. The bimodality follows directly from the distribution of $m_\mathrm{above}$ shown in Fig.\ref{fig:last_aboveHM_vcut30}.} 
    \label{fig:cdmlumifunc_above}
\end{figure}

Fig.~\ref{fig:cdmlumifunc_above} shows the stellar mass function of ghost galaxies for our fiducial $V_\mathrm{cut}=30\,\kms$ model, assuming that 1\% of the available baryonic mass at $z_\mathrm{above}$ is converted to long-lived stars, i.e. $M_{\star} = 0.01 \times (\Omega_{b}/\Omega_{m}) m_\mathrm{above}$. As we discuss below, although this is likely to be a significant overestimate, it shows the overall behavior clearly. The bimodality of the stellar mass function follows directly from the bimodality of $m_\mathrm{above}$ \reply{shown in Fig.~\ref{fig:last_aboveHM_vcut30}} (note that the mass bins in the two figures are different), with the higher-mass peak corresponding to merger tree branches that fall below the cooling threshold for the last time \textit{after} reionization. Lines of different color show the contributions from different bins of present-day virial mass. Lower halo mass bins contribute galaxies of lower stellar mass, reflecting the requirement that the minor branch of the tree in which the ghost forms must exceed the cooling threshold. The range of stellar masses increases with present-day mass, reflecting a wider range of threshold crossing times.

The simplistic assumption of a high and constant star formation efficiency therefore results in a population of ghost galaxies with a peak cosmic abundance of one object with mass $2.5 \lesssim M_{\star} \lesssim 4.5 \times 10^{6}\,\mathrm{M_{\odot}}$ (comparable to the mass of a classical Milky Way satellite) per $5^{3}\,\mathrm{Mpc^{3}}$ volume ($h=0.71$), and a significantly higher density of objects with masses comparable to the ultrafaint Milky Way satellites. The abundance of the most massive ghosts is therefore (in this optimistic estimate) similar to that of Milky Way-mass dark matter halos ($\sim10^{12}\,\mathrm{M_{\odot}}$, $\sim10^{-2}\,\mathrm{Mpc^{-3}\,dex^{-1}}$). Although the conditions that give rise to ghost galaxies are rare, this is compensated by the fact that halos in the relevant mass range are relatively numerous. Of course, `ordinary' dwarfs with stellar masses in the range of Fig.~\ref{fig:cdmlumifunc_above} may be more than an order of magnitude more numerous than this (likely overestimated) prediction for ghost galaxies. In the stellar mass range $\sim10^{9}< M_{\star} < 10^{11}\,\mathrm{M_{\odot}}$, the luminosity function of field galaxies is approximately constant 
 at $\sim10^{-2}\,\mathrm{Mpc^{-3}\,dex^{-1}}$; it then increases by approximately an order of magnitude as $M_{\star}$ decreases from $\sim10^{9} \mathrm{M_{\odot}}$ to $\sim10^{7}\,\mathrm{M_{\odot}}$ \citep[e.g.][]{Wright:2017aa,Bullock:2017aa}. It may be possible, however, to distinguish ghosts as outliers in the plane of magnitude and size or surface brightness, as we discuss in the following section. 

\begin{figure}
    \includegraphics[width=\columnwidth]{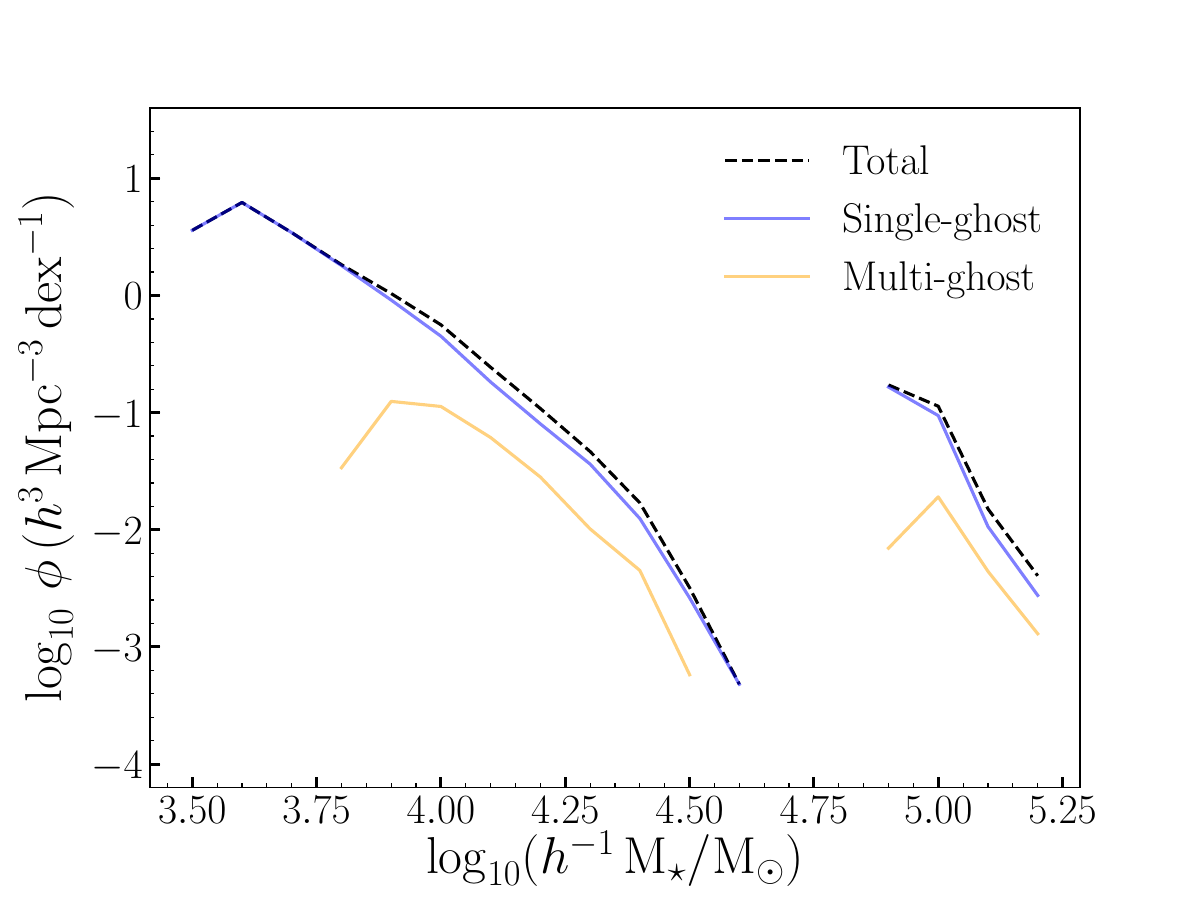}
    \caption{The $z=0$ stellar mass function of ghostly galaxies (dashed black line) for our $V_\mathrm{cut}=30\mathrm{km\,s^{-1}}$ model assuming the SMHM relation of \citet{Behroozi_2019}. Solid lines separate the contributions from galaxies comprising only one ghost (blue) and those formed by the merging of multiple ghosts (orange). The typical ghost stellar mass is reduced compared to the assumption of a constant galaxy formation efficiency in Fig.\ref{fig:cdmlumifunc_above}.}
    \label{fig:Behvcut30lumi}
\end{figure}

Fig.~\ref{fig:Behvcut30lumi} shows a somewhat more realistic estimate based on the mass- and redshift-dependent SMHM relations of \citet{Behroozi_2019}, again for our $V_\mathrm{cut}=30\,\kms$ model. Compared to Fig.~\ref{fig:cdmlumifunc_above}, the stellar mass range is shifted to lower masses by 1.5~dex, and the higher mass peak is significantly narrower. The abundance of galaxies in this peak is similar to the previous estimate. These differences simply reflect the lower star formation efficiency inferred by \citet{Behroozi_2019} compared to the simple assumption of 1\% of available baryons. The amplitude of the \citet{Behroozi_2019} SMHM relation at $M_\mathrm{vir}\sim10^{8}\,\mathrm{M_{\odot}}$ decreases by two orders of magnitude from $z=10$ to $z=1$. This reduces the stellar masses in more massive ghost galaxy halos (which cross the threshold at lower redshift on average) by a larger factor, relative to Fig.~\ref{fig:cdmlumifunc_above}. The steepening slope of the \citet{Behroozi_2019} SMHM relations over the same range of redshift also contributes to the narrower high-mass peak. According to this prediction, the most massive and hence readily detectable ghosts have stellar masses $\approx10^{5}\,\mathrm{M_{\odot}}$. This is slightly less massive than the faintest classical Milky Way satellites, such as Draco and Ursa Minor, $M_{V}\approx-8$ \citep{McConnachie_local_group_2021}, and comparable to the ultrafaints or brighter globular clusters.

\begin{figure}
    \centering
    \includegraphics[width=\columnwidth]{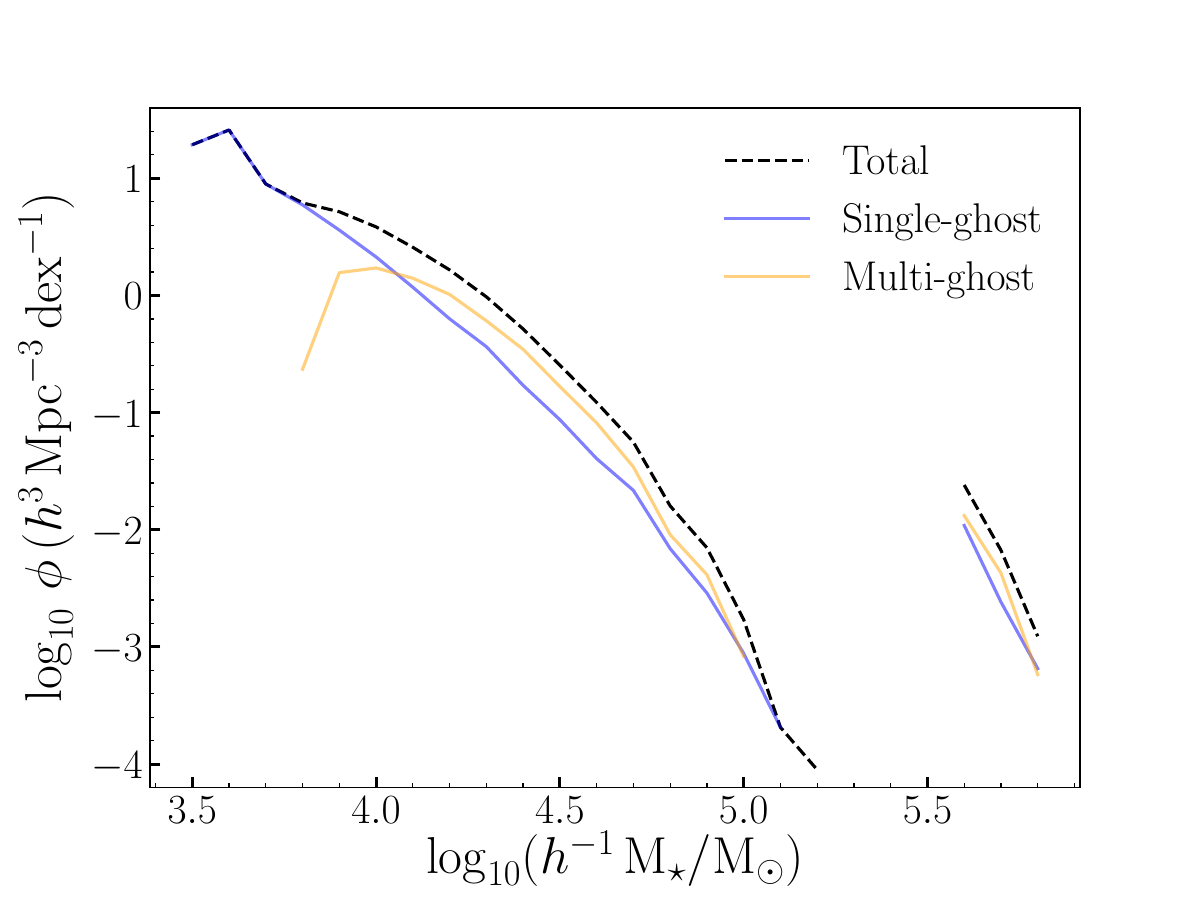}
    \caption{The $z=0$ stellar mass function of ghostly galaxies predicted using the \citet{Behroozi_2019} SMHM relation, as 
 in Fig.\ref{fig:Behvcut30lumi}, but here for a model assuming stronger IGM heating due to reionization, $V_\mathrm{cut} = 40\,\mathrm{km\,s^{-1}}$. Again, solid lines separate the contributions from single (blue) and multiple (orange) ghost trees, and the total is shown by the black dashed line. The typical stellar masses of ghosts are higher than in Fig.~\ref{fig:Behvcut30lumi}, and the number of trees with multiple ghosts exceeds the number of trees with a single ghost at most masses.}
    \label{fig:Behvcut40lumi}
\end{figure}

In the $V_\mathrm{cut}=30\,\kms$ model, the ghost population is dominated by trees producing only a single ghost. Fig.~\ref{fig:Behvcut40lumi} shows the mass function for our $V_\mathrm{cut}=40\,\kms$ model, in which the number of trees producing multiple ghosts is slightly greater than that of trees producing only one (except at the very lowest masses). This suggests that stronger effective reionization results not only in more massive ghost galaxies but also, potentially, a larger fraction with multiple structural components and stellar populations.

\subsection{Merger mass ratios and times}
\label{sec:merger_mass_ratios_and_times}

\begin{figure*}
    \includegraphics[width=\linewidth, trim=0cm 0 0 0, clip=True]{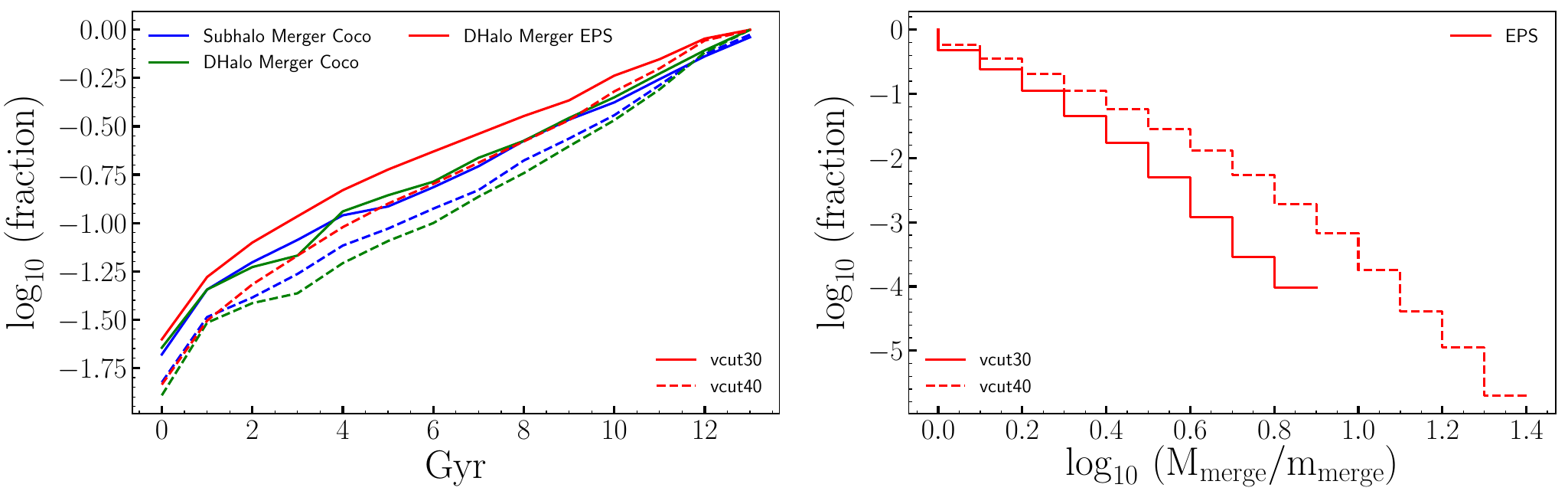}
    \caption{Left: the cumulative fraction of ghosts merging with the main branch later than a given look-back time. Solid and dashed lines refer to our $V_\mathrm{cut}=30$ and $40\,\mathrm{km\,s^{-1}}$ models, respectively. We apply weights to the EPS to reflect the true number counts of an $N$-body simulation and divide each time bin by the total number of ghosts. The red and blue lines correspond to mergers between virialized halos in the EPS and \coco{} trees, respectively. The green lines correspond to the merger of the subhalo associated with the merger tree branch of the ghost galaxy in the \coco{} trees, which may be several snapshots after the virialized halos merge. Right: the cumulative fraction  of mergers between the ghost branch and the main branch of the merger tree with a progenitor halo mass ratio larger than a given value, measured at the time immediately before those two branches merge. Solid and dashed lines again correspond to $V_\mathrm{cut}=30$ and $40\,\mathrm{km\,s^{-1}}$.}
    \label{fig:merger_time_mass}
\end{figure*}

Fig.~\ref{fig:merger_time_mass} shows the distribution of halo mass ratios ($M_\mathrm{merge}/m_\mathrm{merge}$, see Sec.~\ref{sec:minor_branch_and_defns}) and look-back times (corresponding to $z_\mathrm{above}$) for ghost galaxy progenitors merging into their dark main branches in our EPS trees (red lines). This gives an impression of the likely degree of  similarity between the dynamics of ghost galaxies and `ordinary' dwarf galaxies. We find that half of all mergers between ghost galaxies and their dark main branches in \coco{} occur before $z=9$, and almost all are near equal-mass mergers, with only 10\% involving a halo mass ratio greater than 2:1. 

This suggests that they may not be significantly more extended than their counterparts with star-forming main branches of similar final mass \citep[e.g.][]{Amorisco:2017aa_stellar_halos}. Only that fraction of progenitors with high mass ratios are likely to be much more diffuse (stellar halo-like). Higher mass ratios naturally correlate with later mergers. A model with $V_\mathrm{cut}=40\,\mathrm{km\,s^{-1}}$ (dashed lines) results in a slightly larger fraction of high mass ratio mergers at relatively higher redshift. We provide another simple estimate of the dynamical similarity of ghosts and normal dwarf galaxies in the next section.

Our $N$-body merger trees show a similar distribution of merger times defined in an analogous way (mergers between independent halos as defined by the DHalo algorithm), perhaps with a slight tendency toward fewer late mergers. The  $N$-body trees allow for the the ghost progenitor to be tracked as a subhalo after $z_\mathrm{merge}$. We find that the distribution of merger times for these subhalos is not significantly different from that of the DHalos, likely because the mass ratios are low and the halos involved are relatively close to the resolution limit to start with.\footnote{In \coco{}, a halo of $10^{8}\,\mathrm{M_\odot}$ is resolved with $\sim600$ particles and will therefore fall below the $\sim10$-particle halo-finding limit when it has been stripped to $\sim2\%$ of its initial mass.} We do not show the merger mass ratio distribution from the $N$-body simulation because we find that the partition of mass between the two halos in the time step(s) immediately before the merger suffers from what appears to be a systematic effect of the halo-finding algorithm, such that the mass of the minor branch at that time is often underestimated (this can be seen in Figure~\ref{fig:finalHM_hist_figure}). Overall, the $N$-body trees support our conclusions based on the EPS trees. 

\subsection{Density profiles}
\label{sec:profiles}

\begin{figure}
    \includegraphics[width=\linewidth, trim=0cm 0 0 0, clip=True]{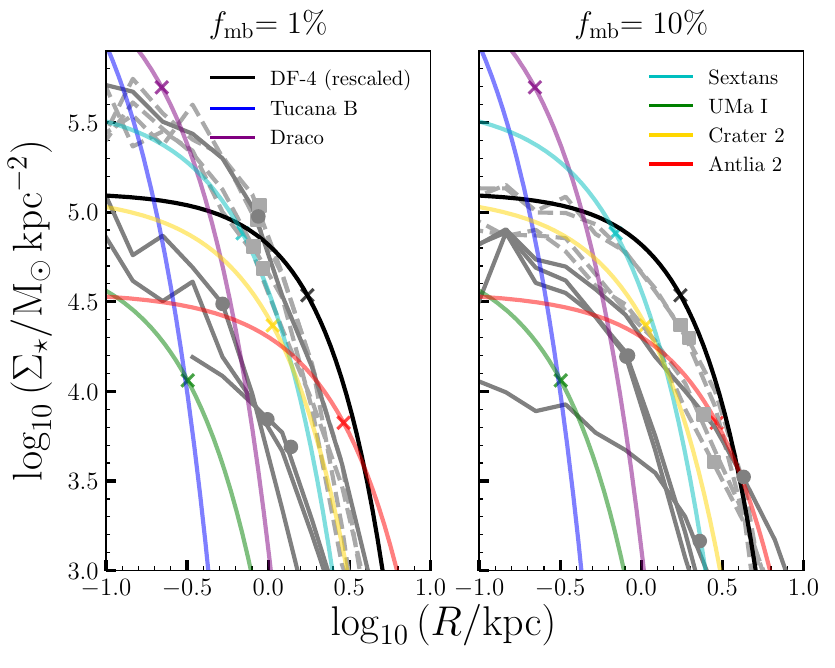}
    \caption{Stellar mass surface density profiles for the most massive ghost galaxies, predicted with our particle tagging prescription (gray lines) for $V_\mathrm{cut}=30\,\mathrm{km\,s^{-1}}$ (solid) and $V_\mathrm{cut}=40\,\mathrm{km\,s^{-1}}$  (dashed). Gray circles and squares mark the respective half-light radii of these profiles. Left and right panels show results with particle tagging most bound faction parameters of $1\%$ and $10\%$ respectively. In both panels, the colored lines show profiles for a selection of real LSB dwarf galaxies as given in the legend; a description of these data and references are given in the text. The crosses mark the half-light radii of the dwarfs. The half-light radius of Tucana B is $\sim80$~pc, outside the range of radii shown.}
    \label{fig:density}
\end{figure}

\begin{figure}
    \includegraphics[width=\linewidth, trim=0cm 0 0 0, clip=True]{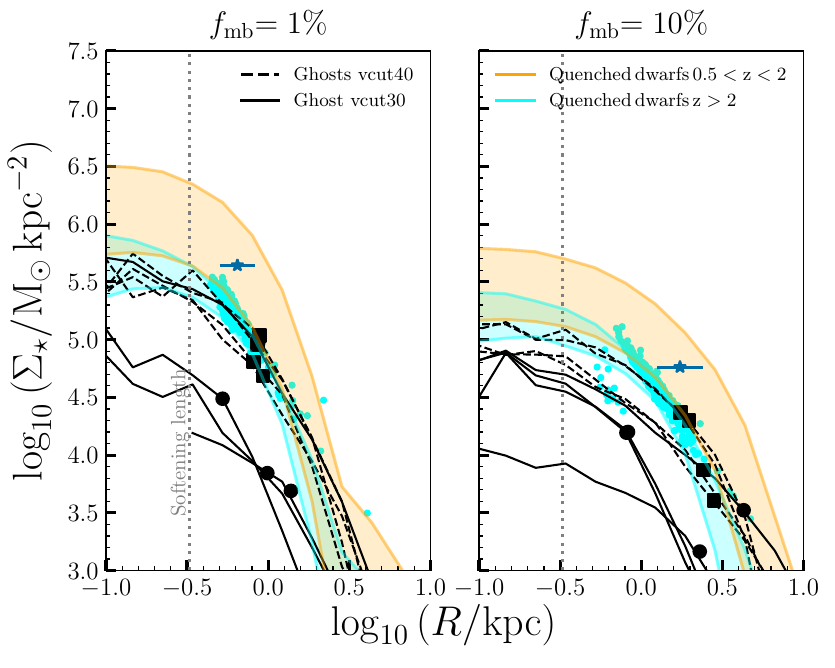}
    \caption{A comparison of stellar mass surface density profiles for massive ghost galaxies (black solid and dashed lines, repeated from Fig.~\ref{fig:density}) and isolated `ordinary' dwarfs that are `quenched' as the result of falling below the virial mass threshold for cooling at $z>0.5$. Shaded regions show the $10\%$--$90\%$ range of profiles for quenched dwarfs crossing the cooling threshold at $0.5 < z < 2$ (orange) and $z>2$ (blue). Half-mass radii for individual dwarfs in the  $z>2$ sample are shown by blue squares. For the $0.5 < z < 2$, we show only the mean ($\pm1\,\sigma$ half-light radius (dark-blue star). The softening length of the \coco{} simulation is indicated by the vertical dotted line.}
    \label{fig:density_ghost_quenchedDwarf}
\end{figure}

We now estimate the stellar mass surface density profiles of brighter ghost galaxies using a technique similar to that of \citet[][D21]{StellarHalo_Deason_2021}. D21 explored the stellar mass accretion histories of present-day dwarfs assuming four different models of the SMHM relation and mass thresholds for galaxy formation. Their CDM model A1 corresponds to a similar set of assumptions to the $V_\mathrm{cut} = 30\,\mathrm{km\,s^{-1}}$ model we use to predict stellar masses (in our case based on the \citealt{Behroozi_2019} SMHM relation). D21 used a simplified `particle tagging' procedure to predict the density profiles of the accreted stellar halos of dwarf galaxies in halos of different present-day mass. These stellar halos are the analogs of our ghost galaxies, for systems in which star formation occurs in the main branch (i.e. the vast majority of dwarf galaxies). The results of D21 therefore already provide some insight into the likely structural properties of ghost galaxies. As described in detail by \citet{Amorisco:2017aa_stellar_halos}, mergers with low mass ratios (total mass, since stars are dynamically insignificant in these systems) produce remnants that have similar structure to their progenitors. Extended halo components are thus built mainly through a succession of lower mass ratio mergers. A steeply falling SMHM relation then necessarily implies that dwarf stellar halos are extremely faint. D21 conclude that they are likely undetectable ($\lesssim30\,\mathrm{mag\,arcsec^{-2}}$) even in stacks of $\sim100$ systems in the full-depth Rubin LSST survey \citep{Ivezic:2019aa} unless stars are able to form efficiently in significantly lower mass halos than expected from the standard cooling threshold arguments (section~\ref{sec:define_ghosts}).

We have carried out a similar experiment to predict the surface brightness profiles of ghost galaxies, using the stellar mass estimates discussed in the previous section together with a simplified particle tagging technique applied to merger trees that we identify with ghosts in the \coco{} simulation. We first identify $z_\mathrm{above}$ as the characteristic time for star formation and then simply select a fixed fraction, $f_\mathrm{mb}$, of the most bound dark matter particles in the ghost galaxy branch halo at this time, in rank order of binding energy. For example, taking $f_\mathrm{mb}=1\%$, we select the top 1\% most bound particles in the halo. We distribute the stellar mass given by the \citet{Behroozi_2019} relation uniformly among these particles. We then recover a surface mass density profile for the ghost galaxy at $z=0$ from the distribution of its tagged particles. 

A `continuous' particle tagging approach would allow for diffusion in the orbits of stars formed at different times \citep[e.g.][]{Le-Bret:2015aa,Cooper:2017aa}. However, we cannot use this approach because our stellar mass estimates are based on the halo mass at a single point in time, $z_\mathrm{above}$. Instead, we tag all of the stellar mass to the halo at $z_\mathrm{above}$, effectively assuming all star formation to occur at this time. The constraints on an appropriate $f_\mathrm{mb}$ are very weak \citep[see][for a detailed discussion]{Cooper:2017aa}. We therefore make two sets of predictions for $f_\mathrm{mb} = 1\%$  and $10\%$, to span a plausible range of possibilities. Lower $f_\mathrm{mb} = 1$ values produce more concentrated initial density profiles.

Fig.~\ref{fig:density} shows the resulting stellar mass surface density profiles of the most massive ghost galaxies in our $V_\mathrm{cut} = 30$ (solid) and $40\,\mathrm{km\,s^{-1}}$ (dashed) models. We compare these to profile shapes and scale lengths for a variety of observed dwarf galaxies. We restrict our comparison to the most massive ghost galaxies in our model because they are most likely to be observable beyond the Local Group and also because the half-mass radii of less massive ghosts approach the spatial resolution limit of \coco{} ($230\,h^{-1}\,\mathrm{pc}$). We note that all our ghost galaxies are isolated at $z=0$, in the sense that they are identified as independent systems by the \coco{} halo-finding algorithm, although we do not track their individual interaction histories to check whether they were satellites at earlier times, nor do we examine their larger-scale environment. However, our sample of observed systems for comparison includes both isolated dwarfs and satellites, as follows.

Tucana B is an isolated dwarf in the Local Group with an `ultra faint' luminosity of $\sim 5\times10^{4}\,\mathrm{L_{\odot}}$ and a half-light radius $80\pm40\,\mathrm{pc}$, lacking recent star formation \citep{sand:2022aa_tucb}. Tuc B provides a useful point of reference for an isolated, early-forming `reionization fossil' in a mass range comparable to our predictions for the most massive ghost galaxies, potentially free from effects due to interactions with a more massive galaxy. In this respect, it is notably compact, with a significantly more concentrated profile than ghosts of a similar or greater mass, regardless of our choice of $f_\mathrm{mb}$. 

Dragonfly 4 is a well-studied UDG that has been claimed to be deficient in dark matter \citep{PvD2019_2ndMissingDM}. \citet{Montes:2020aa_df4} report the surface brightness profile of this object to $\approx20\,\mathrm{mag\,arcsec^{-2}}$. They measure a stellar mass of $3.6\times10^{7}\,\mathrm{M_{\odot}}$, greater than our ghosts; since we are interested here only in the shape of the surface brightness profile, we simply scale down the amplitude of the profile reported by \citealt{Montes:2020aa_df4} by a factor of 10 in Fig.~\ref{fig:density}. \citealt{Montes:2020aa_df4} also find evidence that the galaxy is tidally distorted by interaction with a more massive neighbor.

Crater 2 \citep{Torrealba:2016ab} has the fourth-largest half-light radius among the satellites of the Milky Way, $1.1\,\mathrm{kpc}$, but a luminosity of only $M_{V}\approx-8$. Consequently, it has an extremely low surface brightness of $\sim \,30\,\mathrm{mag\,arcsec^{-2}}$. It also has an unusually low velocity dispersion, $\sigma_{\mathrm{los}}\sim2.7\,\mathrm{km\,s^{-1}}$ \citep{Caldwell:2016aa}. The observed stellar mass suggests a significantly more massive halo, $V_{c}\sim20$--$30\,\mathrm{km\,s^{-1}}$. Although substantial tidal stripping would explain its low $\sigma_{\mathrm{los}}$, this explanation is in tension with its large size \citep{Borukhovetskaya:2022aa}. Antlia 2 \citep{Torrealba_2019_Antlia2} is another Milky Way companion with an exceptionally large half light radius of $2.9\,\mathrm{kpc}$ and a velocity dispersion of $\sigma_{\mathrm{los}}\sim5.7\,\mathrm{km\,s^{-1}}$. Like Crater 2, it is as yet unclear whether tidal disruption is a sufficient or unique explanation for these properties \citep[e.g.][]{ji_2021_antlia_crater}. 

We also show profiles for three of the classical Milky Way satellites with comparable luminosity to the brightest ghost galaxies in our model, Draco, Sextans, and UMa I, using data from \citet{McConnachie_local_group_2021}. All three have relatively large half light radii compared to the average for satellites of their stellar mass. Of these, Sextans ($M_{\star}\approx 5\times10^{5}\,\mathrm{M_{\odot}}$) has the most similar density distribution to our predictions for ghost galaxies. It shows signs of density and velocity substructure that may be evidence of accretion \citep{Cicuendez:2018aa}. Other dSph satellites, even the more extended examples shown here, appear significantly more compact. 

Finally, Fig.~\ref{fig:density_ghost_quenchedDwarf} compares the predictions for ghost galaxies shown in Fig.~\ref{fig:density} to the average profiles of `ordinary' isolated dwarf galaxies without recent star formation, as predicted by our model. To create this comparison sample, we select main branches that have crossed the cooling threshold in the past but that have fallen below it again at $z \ge 0.5$. This requirement is intended to select a sample of `quenched' dwarfs and hence to exclude (very approximately) those that would be considered young or actively star-forming at the present day. Present-day star-forming dwarfs are not candidate ghost galaxies, by definition. These dwarf galaxies driven to quiescence by the UV background are discussed in detail by \citet{Pereira-Wilson:2023aa}. We compute density profiles in the same way as for the ghosts, in this case estimating a stellar mass at the time at which the \textit{main} branch of the tree is last above the threshold and tagging particles in that branch accordingly. In Fig.~\ref{fig:density_ghost_quenchedDwarf} we further split the `ordinary' quenched dwarfs into those that fall below the cooling threshold at $0.5 < z < 2$ (younger) and $z > 2$ (older). The older subset is of comparable `age' to the majority of ghosts. We find that massive ghosts have stellar masses and surface density profiles broadly similar to those of the older quenched dwarfs. They tend toward the largest sizes (lowest central densities) for that population. Younger quenched dwarfs have central densities an order of magnitude greater than ghosts of similar size. 

\section{Conclusions}
\label{sec:conclusion}

We have explored a `ghost galaxy' scenario for the formation of field dwarf galaxies. Our results are based on a simple `threshold' model of galaxy formation applied to dark matter halo merger trees constructed using  the EPS method \citep{Parkinson2007} and the \coco{} $N$-body simulation \citep{Hellwing_2016}. Star formation is inhibited in halos with virial temperatures below the cooling limit of atomic hydrogen and, at redshifts lower than $z=z_\mathrm{reion}$, in halos with virial velocity $V_\mathrm{vir} < V_\mathrm{cut}$. We find that the halo mass range, overall number, and luminosity function of ghost galaxies are sensitive to the suppressive effect of the UV background. We have examined models with $V_\mathrm{cut} = 30\,\kms$ (`weaker' reionization) and $40\,\kms$ (`stronger' reionization). Our specific results are as follows:

\begin{itemize}
    \item Ghost galaxies form in a halo mass range approximately $2 \times 10^{9} h^{-1} \,\mathrm{M_{\odot}} < M_{\mathrm{vir}} < 1 \times 10^{10}\,h^{-1}\,\mathrm{M_{\odot}}$ with $V_\mathrm{cut}=30\,\kms$ or $2 \times 10^{9}\,h^{-1} \,\mathrm{M_{\odot}} < M_{\mathrm{vir}} < 2 \times 10^{10}\,\,h^{-1}\,\mathrm{M_{\odot}}$ with $V_\mathrm{cut}=40\,\kms$ (Fig.~\ref{fig:abovefrac_vcut3040}).
    \item For $V_\mathrm{cut}=30\,\kms$, ghost galaxies are most likely to occur in halos with $M_{\mathrm{vir}}\simeq4\times10^{9}\,h^{-1}\,\mathrm{M_{\odot}}$ ($\approx5$~per~cent of all halos of that mass). For $V_\mathrm{cut}=40\,\kms$, they are most likely at $M_{\mathrm{vir}} \simeq 8\times10^{9} h^{-1}\,\mathrm{M_{\odot}}$ ($\approx20$~per~cent of all halos of that mass; Fig.~\ref{fig:abovefrac_cocoEPS}).
    \item These characteristic masses and occupation fractions vary in a non-trivial way when the redshift of reionization and the cooling threshold before reionization are adjusted within plausible bounds. With the typical mass accretion rates of halos fixed, given $V_\mathrm{cut}$, the requirement that the main branch must remain below the cooling threshold but at least one minor branch must exceed it leads to a maximum in the fraction (and overall number) of ghosts for a particular $z_\mathrm{reion}$.
    \item By assigning stellar mass to halos following the prescriptions of \citet{Behroozi_2019}, we predict ghost galaxies have a bimodal luminosity distribution: an `ultra faint` population that accounts for the majority of systems and a smaller but significant population of brighter objects (Figs.~\ref{fig:Behvcut30lumi} and \ref{fig:Behvcut40lumi}). Analogous to the typical satellite luminosity function of Milky Way--like galaxies, these populations correspond to systems forming stars before and after reionization, respectively (Fig.~\ref{fig:last_aboveHM_vcut30}.)
    \item The brighter ghost galaxy population has a characteristic stellar mass $\gtrsim 10^{5}\,h^{-1}\,\msol$ ($V_\mathrm{cut} = 30\kms$), \reply{comparable to the `UDG-like' satellites of the Milky Way, Crater 2 ($M_{\star} \simeq 10^{5.55}\,\msol$) and Antlia 2 ($M_{\star} \simeq 10^{6.22}\,\msol$) \citep{ji_2021_antlia_crater}, and somewhat lower than estimates for the diffuse M31 satellites And II \citep[$7.6\times10^6\,\msol$l][]{McConnachie_local_group_2021}, And XIX \citep[$\sim10^6\,\msol$;][]{Collins:2020_andxix}, and And XXXII \citep[$\sim10^7\,\msol$;][]{martin:2013_andxxxii}}. The \reply{characteristic mass} in our model increases by a factor of $\sim3$ for $V_\mathrm{cut} = 40\,\kms$, in which case the most massive ghost galaxies have masses comparable to fainter `classical' Milky Way dwarf satellites such as Sextans and Draco.

    \item \reply{Although the characteristic stellar mass of the brighter ghost galaxy population in our model is similar to the masses of the most diffuse Milky Way satellites, it is substantially below the range of $10^{7}$--$10^{9}\,\msol$ observed for UDGs in the field \citep[e.g.][]{kado-fong:2022_gas_rich_udgs, zaritsky:2023_smudges}.
    The difficulty of producing ghost galaxies with higher masses, given present constraints on reionization and star formation efficiencies at high redshift, is the most significant argument against this scenario providing sufficient explanation for the observed UDG population.}
    
    \item For $V_\mathrm{cut} = 30\kms$, the ghost population consists mostly of systems with a single dominant progenitor; for $V_\mathrm{cut} = 40\kms$, systems with two or more progenitors are equally common (Figs.~\ref{fig:Behvcut30lumi} and \ref{fig:Behvcut40lumi}).
    \item The majority of ghost galaxy progenitors merge with their dark main branches at high redshift, predominantly in mergers that are close to equal mass. A higher value of $V_\mathrm{cut} = 40\kms$ increases the fraction of mergers with higher mass ratios.
    \item We make a simple estimate of the density profiles of the most massive ghost galaxies using a particle tagging prescription in combination with stellar mass estimates from \citet{Behroozi_2019}. We find that the resulting $z=0$ ghost systems have half-light radii comparable to the UDG Dragonfly 4 ($R_{50}\gtrsim 2\,\mathrm{kpc}$ and the unusually faint Milky Way satellite Crater 2, if we assume a larger extent for the stars at the time of their formation ($f_\mathrm{mb}=10\%$). If we assume that the initial extent of the population is relatively compact ($f_\mathrm{mb}=1\%$), the ghosts have  half light radii similar to the classical Milky Way satellite Sextans ($R_{50}\sim 1\,\mathrm{kpc}$). \apc{`Ordinary' dwarf galaxies that form stars at high redshift and later fall below the cooling threshold \citep[e.g][]{Pereira-Wilson:2023aa} have similar sizes to ghosts, reinforcing our conclusion \reply{(based primarily on the predicted mass distribution)} that ghost galaxies are not a unique or sufficient explanation for the ultradiffuse dwarf population.}
\end{itemize}

We conclude that the ghost galaxy mechanism is a plausible, even likely formation scenario for a fraction of faint field dwarf galaxies in $\Lambda$CDM. Such objects, if they exist, would be the most dark-matter-dominated virialized systems that could be probed with stellar kinematics. 

\apc{Our findings are related to those of \citet{Ricotti:2022aa}, who examine models for stellar halos built up around dwarf galaxies by the accretion of ultrafaint `reionization fossil' progenitors. These stellar halos have the same origin as our proposed ghostly galaxies: indeed, they were dubbed `ghostly halos' by \citet{Bovill_Ricotti2011}. The important distinction between our work and that reported in \citet{Ricotti:2022aa} is that we specifically consider the formation of stellar halos in systems without in situ star formation, which have not previously been examined separately. We also include our consideration of the small (but potentially significant) fraction of accreted progenitors that may form stars after reionization.}

\apc{\citet{Ricotti:2022aa} discuss the tentative evidence that stellar halos exist around some of the more massive dwarf galaxies in the Local Group. They collate observations of surface brightness profiles in six isolated Local Group dwarfs that show outer breaks, suggestive of a stellar halo component. As in our work and that of \citet{StellarHalo_Deason_2021}, they find that predictions for the stellar halos of dwarf galaxies are sensitive to the assumed star formation efficiency before reionization. Through comparison of their Local Group data to models, they estimate a star formation efficiency for the progenitors before reionization consistent with extrapolation of the abundance matching relation of \citet{Behroozi:2013aa}. They also find that the density profiles and scale radii of these halos are similar to those of the stars form in situ in the dwarf galaxies.}

\apc{The star-free main branches of ghost galaxies are closely related to the reionization-limited HI clouds (RELHICs) discussed by \citet{Benitez_Llambay_2017} and \citet{Benitez_Llambay_2020}. By definition, RELHICs remain below the star formation threshold but are massive enough to retain baryons against photoevaporation by the UV background. In the simulations studied by \citet{Benitez_Llambay_2017}, RELHICs of mass $M_{200}\sim5\times10^{9}\,\mathrm{M_\odot}$ were found to have baryon fractions $\sim20\%$ of the universal value. Although most of this gas is ionized, the more massive RELHICs support neutral cores. Candidate ghost galaxies could therefore correspond to RELHIC-like systems, which may be detectable in future HI surveys. However, \citet{Benitez_Llambay_2017} found that $\sim50\%$ of all dark matter halos with mass $M_{200}\approx2\times10^{9}\,\mathrm{M_\odot}$ are RELHICs (assuming $z_\mathrm{reion}\approx11$), whereas we predict $\lesssim 5\%$ of halos of this mass host ghost galaxies. Thus, although a large fraction of ghosts may be RELHICs, very few RELHICs are likely to be ghosts.} 

Naively, ghost galaxies might be expected to have very low surface brightness for their mass and hence be a potential contributor to the `ultra diffuse' \reply{dwarf galaxy} population. However, \reply{as noted above,} our results imply that their cosmic abundance is low at \reply{the stellar masses estimated for known UDGs, in particular those outside the Local Group}. \reply{Ghost galaxies (as we define them) therefore appear} unlikely to be the only or even dominant component of \reply{the known UDG} population. \reply{The majority of ghosts have stellar masses so low that they would not be readily detected by current facilities; if they exist, future surveys may be able to observe them in the Local Group \citep[e.g.][]{Newton:2023_hestia_local_group}}. \reply{However,} our simple estimates of their luminosity, size, and dynamical state suggest that ghosts may be hard to distinguish from typical dwarf galaxies, at least under standard assumptions about cosmic reionization.\footnote{\reply{The discrepancy between the upper end of the ghost mass distribution and the masses of known UDGs could be reduced, but is unlikely to be eliminated, by accounting for stronger local reionization or by imposing a less conservative restriction on in situ star formation.}} More detailed quantitative statements about their observability would require explicit simulations of their star formation histories and dynamics. 

We nevertheless find one interesting and general result: the abundance and structure of ghostly galaxies is potentially very sensitive to reionization. At the level of the simple prescription we use here, the strongest constraint on the effective heating of the IGM by the cosmic UV background (parameterized by $V_\mathrm{cut}$ and $z_\mathrm{cut}$) comes from comparison of models to the low-mass end of the Milky Way satellite stellar mass function \citep[e.g.][]{Benson:2002aa,Font:2011aa,Bose:2018aa}. The most appropriate value of $V_\mathrm{cut}$ therefore remains somewhat uncertain, not least because it is unclear which simulated satellite populations should be used for comparison to the Milky Way. More significantly, reionization is expected to occur earlier and to produce a locally stronger suppression of cooling due to the ionizing background (earlier $z_\mathrm{cut}$ and/or higher effective $V_\mathrm{cut}$) in regions of higher density \citep[e.g.][]{Efstathiou:1992aa, Weinmann_2007, Font:2011aa}. In extreme regions, such as the environs of massive galaxy clusters, this local reionization could increase the abundance of ghost galaxies and also increase the disparity between their size and luminosity (the latter reducing with stronger/earlier reionization, the former increasing as in situ star formation is suppressed in more massive host halos). 

Arguing against this, the main branches of cluster progenitors are less likely to satisfy our strict requirement of always remaining below the cooling threshold, because they necessarily collapse earlier than their counterparts in the field \citep[see, e.g.][]{Weinmann_2007}. However, the maximum stellar mass that can form in those branches will always be limited by (local) reionization; given that limitation, relatively massive cluster satellite halos at $z=0$ may have very high accreted stellar mass fractions as the result of mergers with multiple progenitors of similar stellar mass. The abundance of ghostly galaxies in high-density regions may therefore be substantially different from that in the field, particularly if we were to relax our criteria to include dwarfs that are only \textit{dominated} by accreted stars, rather than considering only those formed entirely by accretion. These effects could be explored with more detailed models of local reionization in dense regions.

Although there is evidence that diffuse dwarf galaxies are common in clusters \citep[e.g.][]{Koda_2015_UDGs,Munoz:2015aa,van-der-Burg:2017aa}, it is currently hard to disentangle the enhancement of different modes of dwarf galaxy formation, such as the scenario above, from the effects of higher galaxy density overall in these regions, potential bias toward deeper observations in clusters, and environmental effects that might act on `normal' dwarf galaxies. Upcoming wide-area deep-imaging surveys, including LSST \citep{Ivezic:2019aa}, could address these questions by discovering much larger numbers of very low surface brightness dwarfs and mapping their abundance over larger areas around clusters and in the field.

\begin{acknowledgments}
\reply{The authors thank the anonymous referee for their thoughtful and constructive feedback} and Shaun Cole and John Helly for their assistance with the \citet{Parkinson2007} EPS merger tree code. W.C.W. and A.P.C. are supported by a Yushan Fellowship, awarded to APC by the Taiwan Ministry of Education. APC acknowledges support from Taiwan's National Science and Technology Council under grants 109-2112-M-007-011-MY3 and 112-2112-M-007-017-MY3. This work used high-performance computing facilities operated by the Center for Informatics and Computation in Astronomy (CICA) at National Tsing Hua University. This equipment was funded by the Ministry of Education of Taiwan, the National Science and Technology Council of Taiwan, and National Tsing Hua University. S.B. is supported by the UK Research and Innovation (UKRI) Future Leaders Fellowship (grant number MR/V023381/1). C.S.F. acknowledges support by the European Research Council (ERC) through Advanced Investigator grant DMIDAS (GA 786910). W.A.H. is supported by research grants funded by the National Science Center, Poland, under agreements 2018/31/G/ST9/03388 and 2020/39/B/ST9/03494. This work used the DiRAC@Durham facility managed by the Institute for Computational Cosmology on behalf of the STFC DiRAC HPC Facility (\url{www.dirac.ac.uk}). The equipment was funded by BEIS capital funding via STFC capital grants ST/K00042X/1, ST/P002293/1, ST/R002371/1, and ST/S002502/1; Durham University; and STFC operations grant ST/R000832/1. DiRAC is part of the National e-Infrastructure.

\end{acknowledgments}

%



\software{numpy \citep[][]{Harris2020ArrayNumPy}, matplotlib \citep[][]{Hunter2007Matplotlib:Environment}, astropy \citep[][]{Collaboration2013Astropy:Astronomy,Collaboration2018ThePackage}. Figure 1 was drawn using \url{excalidraw.com}.}

\bibliography{bib}{}

\begin{thebibliography}{}
\expandafter\ifx\csname natexlab\endcsname\relax\def\natexlab#1{#1}\fi
\providecommand{\url}[1]{\href{#1}{#1}}
\providecommand{\dodoi}[1]{doi:~\href{http://doi.org/#1}{\nolinkurl{#1}}}
\providecommand{\doeprint}[1]{\href{http://ascl.net/#1}{\nolinkurl{http://ascl.net/#1}}}
\providecommand{\doarXiv}[1]{\href{https://arxiv.org/abs/#1}{\nolinkurl{https://arxiv.org/abs/#1}}}

\bibitem[{{Amorisco}(2017)}]{Amorisco:2017aa_stellar_halos}
{Amorisco}, N.~C. 2017, \mnras, 464, 2882

\bibitem[{{Amorisco} \& {Loeb}(2016)}]{Amorisco:2016aa}
{Amorisco}, N.~C., \& {Loeb}, A. 2016, \mnras, 459, L51

\bibitem[{{Barbosa} {et~al.}(2020){Barbosa}, {Zaritsky}, {Donnerstein}, {Zhang}, {Dey}, {Mendes de Oliveira}, {Sampedro}, {Molino}, {Costa-Duarte}, {Coelho}, {Cortesi}, {Herpich}, {Hernandez-Jimenez}, {Santos-Silva}, {Pereira}, {Werle}, {Overzier}, {Cid Fernandes}, {Smith Castelli}, {Ribeiro}, {Schoenell}, \& {Kanaan}}]{Barbosa:2020vq}
{Barbosa}, C.~E., {Zaritsky}, D., {Donnerstein}, R., {et~al.} 2020, \apjs, 247, 46

\bibitem[{Behroozi {et~al.}(2019)Behroozi, Wechsler, Hearin, \& Conroy}]{Behroozi_2019}
Behroozi, P., Wechsler, R.~H., Hearin, A.~P., \& Conroy, C. 2019, Monthly Notices of the Royal Astronomical Society, 488, 3143, \dodoi{10.1093/mnras/stz1182}

\bibitem[{{Behroozi} {et~al.}(2013){Behroozi}, {Wechsler}, \& {Conroy}}]{Behroozi:2013aa}
{Behroozi}, P.~S., {Wechsler}, R.~H., \& {Conroy}, C. 2013, \apj, 770, 57

\bibitem[{Benitez-Llambay \& Frenk(2020)}]{Benitez_Llambay_2020}
Benitez-Llambay, A., \& Frenk, C. 2020, Monthly Notices of the Royal Astronomical Society, 498, 4887, \dodoi{10.1093/mnras/staa2698}

\bibitem[{{Ben{\'\i}tez-Llambay} {et~al.}(2017){Ben{\'\i}tez-Llambay}, {Navarro}, {Frenk}, {Sawala}, {Oman}, {Fattahi}, {Schaller}, {Schaye}, {Crain}, \& {Theuns}}]{Benitez_Llambay_2017}
{Ben{\'\i}tez-Llambay}, A., {Navarro}, J.~F., {Frenk}, C.~S., {et~al.} 2017, \mnras, 465, 3913, \dodoi{10.1093/mnras/stw2982}

\bibitem[{{Benson}(2010)}]{Benson:2010aa}
{Benson}, A.~J. 2010, \physrep, 495, 33

\bibitem[{{Benson} {et~al.}(2003){Benson}, {Bower}, {Frenk}, {Lacey}, {Baugh}, \& {Cole}}]{Benson:2003aa}
{Benson}, A.~J., {Bower}, R.~G., {Frenk}, C.~S., {et~al.} 2003, \apj, 599, 38, \dodoi{10.1086/379160}

\bibitem[{{Benson} {et~al.}(2002{\natexlab{a}}){Benson}, {Frenk}, {Lacey}, {Baugh}, \& {Cole}}]{Benson:2002aa}
{Benson}, A.~J., {Frenk}, C.~S., {Lacey}, C.~G., {Baugh}, C.~M., \& {Cole}, S. 2002{\natexlab{a}}, \mnras, 333, 177

\bibitem[{{Benson} {et~al.}(2002{\natexlab{b}}){Benson}, {Lacey}, {Baugh}, {Cole}, \& {Frenk}}]{Benson:2002paper1}
{Benson}, A.~J., {Lacey}, C.~G., {Baugh}, C.~M., {Cole}, S., \& {Frenk}, C.~S. 2002{\natexlab{b}}, \mnras, 333, 156, \dodoi{10.1046/j.1365-8711.2002.05387.x}

\bibitem[{Benson {et~al.}(2019)Benson, Ludlow, \& Cole}]{Benson_2019}
Benson, A.~J., Ludlow, A., \& Cole, S. 2019, Monthly Notices of the Royal Astronomical Society, 485, 5010, \dodoi{10.1093/mnras/stz695}

\bibitem[{{Borukhovetskaya} {et~al.}(2022){Borukhovetskaya}, {Navarro}, {Errani}, \& {Fattahi}}]{Borukhovetskaya:2022aa}
{Borukhovetskaya}, A., {Navarro}, J.~F., {Errani}, R., \& {Fattahi}, A. 2022, \mnras, 512, 5247

\bibitem[{{Bose} {et~al.}(2018){Bose}, {Deason}, \& {Frenk}}]{Bose:2018aa}
{Bose}, S., {Deason}, A.~J., \& {Frenk}, C.~S. 2018, \apj, 863, 123

\bibitem[{{Bovill} \& {Ricotti}(2009)}]{Bovill:2009aa}
{Bovill}, M.~S., \& {Ricotti}, M. 2009, \apj, 693, 1859

\bibitem[{{Bovill} \& {Ricotti}(2011)}]{Bovill_Ricotti2011}
---. 2011, \apj, 741, 17, \dodoi{10.1088/0004-637X/741/1/17}

\bibitem[{{Bower} {et~al.}(2006){Bower}, {Benson}, {Malbon}, {Helly}, {Frenk}, {Baugh}, {Cole}, \& {Lacey}}]{Bower:2006aa}
{Bower}, R.~G., {Benson}, A.~J., {Malbon}, R., {et~al.} 2006, \mnras, 370, 645

\bibitem[{{Bullock} \& {Boylan-Kolchin}(2017)}]{Bullock:2017aa}
{Bullock}, J.~S., \& {Boylan-Kolchin}, M. 2017, \araa, 55, 343

\bibitem[{{Bullock} {et~al.}(2000){Bullock}, {Kravtsov}, \& {Weinberg}}]{BKW}
{Bullock}, J.~S., {Kravtsov}, A.~V., \& {Weinberg}, D.~H. 2000, \apj, 539, 517, \dodoi{10.1086/309279}

\bibitem[{{Caldwell} {et~al.}(2017){Caldwell}, {Walker}, {Mateo}, {Olszewski}, {Koposov}, {Belokurov}, {Torrealba}, {Geringer-Sameth}, \& {Johnson}}]{Caldwell:2016aa}
{Caldwell}, N., {Walker}, M.~G., {Mateo}, M., {et~al.} 2017, \apj, 839, 20, \dodoi{10.3847/1538-4357/aa688e}

\bibitem[{{Cicu{\'e}ndez} \& {Battaglia}(2018)}]{Cicuendez:2018aa}
{Cicu{\'e}ndez}, L., \& {Battaglia}, G. 2018, \mnras, 480, 251, \dodoi{10.1093/mnras/sty1748}

\bibitem[{{Cole} {et~al.}(2000){Cole}, {Lacey}, {Baugh}, \& {Frenk}}]{Cole_Lacey_2000_Galform}
{Cole}, S., {Lacey}, C.~G., {Baugh}, C.~M., \& {Frenk}, C.~S. 2000, \mnras, 319, 168, \dodoi{10.1046/j.1365-8711.2000.03879.x}

\bibitem[{{Collins} {et~al.}(2020){Collins}, {Tollerud}, {Rich}, {Ibata}, {Martin}, {Chapman}, {Gilbert}, \& {Preston}}]{Collins:2020_andxix}
{Collins}, M. L.~M., {Tollerud}, E.~J., {Rich}, R.~M., {et~al.} 2020, \mnras, 491, 3496, \dodoi{10.1093/mnras/stz3252}

\bibitem[{{Couchman} \& {Rees}(1986)}]{Couchman:1986aa}
{Couchman}, H.~M.~P., \& {Rees}, M.~J. 1986, \mnras, 221, 53

\bibitem[{Deason {et~al.}(2021)Deason, Bose, Fattahi, Amorisco, Hellwing, \& Frenk}]{StellarHalo_Deason_2021}
Deason, A.~J., Bose, S., Fattahi, A., {et~al.} 2021, Monthly Notices of the Royal Astronomical Society, 511, 4044, \dodoi{10.1093/mnras/stab3524}

\bibitem[{{Efstathiou}(1992)}]{Efstathiou:1992aa}
{Efstathiou}, G. 1992, \mnras, 256, 43P

\bibitem[{{Fielder} {et~al.}(2023){Fielder}, {Jones}, {Sand}, {Bennet}, {Crnojevi{\'c}}, {Karunakaran}, {Mutlu-Pakdil}, \& {Spekkens}}]{fielder2023}
{Fielder}, C.~E., {Jones}, M.~G., {Sand}, D.~J., {et~al.} 2023, \apjl, 954, L39, \dodoi{10.3847/2041-8213/acf0c3}

\bibitem[{{Font} {et~al.}(2011){Font}, {Benson}, {Bower}, {Frenk}, {Cooper}, {De Lucia}, {Helly}, {Helmi}, {Li}, {McCarthy}, {Navarro}, {Springel}, {Starkenburg}, {Wang}, \& {White}}]{Font:2011aa}
{Font}, A.~S., {Benson}, A.~J., {Bower}, R.~G., {et~al.} 2011, \mnras, 417, 1260

\bibitem[{{Gnedin}(2000)}]{Gnedin:2000aa}
{Gnedin}, N.~Y. 2000, \apj, 542, 535

\bibitem[{{Harris} {et~al.}(2020){Harris}, {Millman}, {van der Walt}, {Gommers}, {Virtanen}, {Cournapeau}, {Wieser}, {Taylor}, {Berg}, {Smith}, {Kern}, {Picus}, {Hoyer}, {van Kerkwijk}, {Brett}, {Haldane}, {del R{\'\i}o}, {Wiebe}, {Peterson}, {G{\'e}rard-Marchant}, {Sheppard}, {Reddy}, {Weckesser}, {Abbasi}, {Gohlke}, \& {Oliphant}}]{Harris2020ArrayNumPy}
{Harris}, C.~R., {Millman}, K.~J., {van der Walt}, S.~J., {et~al.} 2020, \nat, 585, 357, \dodoi{10.1038/s41586-020-2649-2}

\bibitem[{{Hellwing} {et~al.}(2021){Hellwing}, {Cautun}, {van de Weygaert}, \& {Jones}}]{hellwing_coco_environment}
{Hellwing}, W.~A., {Cautun}, M., {van de Weygaert}, R., \& {Jones}, B.~T. 2021, \prd, 103, 063517, \dodoi{10.1103/PhysRevD.103.063517}

\bibitem[{Hellwing {et~al.}(2016)Hellwing, Frenk, Cautun, Bose, Helly, Jenkins, Sawala, \& Cytowski}]{Hellwing_2016}
Hellwing, W.~A., Frenk, C.~S., Cautun, M., {et~al.} 2016, Monthly Notices of the Royal Astronomical Society, 457, 3492, \dodoi{10.1093/mnras/stw214}

\bibitem[{{Hoeft} {et~al.}(2006){Hoeft}, {Yepes}, {Gottl{\"o}ber}, \& {Springel}}]{Hoeft:2006aa}
{Hoeft}, M., {Yepes}, G., {Gottl{\"o}ber}, S., \& {Springel}, V. 2006, \mnras, 371, 401

\bibitem[{Hunter(2007)}]{Hunter2007Matplotlib:Environment}
Hunter, J.~D. 2007, Computing in Science Engineering, 9, 90, \dodoi{10.1109/MCSE.2007.55}

\bibitem[{{Ikeuchi}(1986)}]{Ikeuchi:1986aa}
{Ikeuchi}, S. 1986, \apss, 118, 509

\bibitem[{{Ivezi{\'c}} {et~al.}(2019){Ivezi{\'c}}, {Kahn}, {Tyson}, {Abel}, {Acosta}, {Allsman}, {Alonso}, {AlSayyad}, {Anderson}, {Andrew}, \& et~al.}]{Ivezic:2019aa}
{Ivezi{\'c}}, {\v{Z}}., {Kahn}, S.~M., {Tyson}, J.~A., {et~al.} 2019, \apj, 873, 111

\bibitem[{{Ji} {et~al.}(2021){Ji}, {Koposov}, {Li}, {Erkal}, {Pace}, {Simon}, {Belokurov}, {Cullinane}, {Da Costa}, {Kuehn}, {Lewis}, {Mackey}, {Shipp}, {Simpson}, {Zucker}, {Hansen}, {Bland-Hawthorn}, \& {S5 Collaboration}}]{ji_2021_antlia_crater}
{Ji}, A.~P., {Koposov}, S.~E., {Li}, T.~S., {et~al.} 2021, \apj, 921, 32, \dodoi{10.3847/1538-4357/ac1869}

\bibitem[{{Jiang} {et~al.}(2019){Jiang}, {Dekel}, {Freundlich}, {Romanowsky}, {Dutton}, {Macci{\`o}}, \& {Di Cintio}}]{jiang2019}
{Jiang}, F., {Dekel}, A., {Freundlich}, J., {et~al.} 2019, \mnras, 487, 5272, \dodoi{10.1093/mnras/stz1499}

\bibitem[{{Jiang} {et~al.}(2014){Jiang}, {Helly}, {Cole}, \& {Frenk}}]{jiang2014_dhalos}
{Jiang}, L., {Helly}, J.~C., {Cole}, S., \& {Frenk}, C.~S. 2014, \mnras, 440, 2115, \dodoi{10.1093/mnras/stu390}

\bibitem[{{Jones} {et~al.}(2021){Jones}, {Bennet}, {Mutlu-Pakdil}, {Sand}, {Spekkens}, {Crnojevi{\'c}}, {Karunakaran}, \& {Zaritsky}}]{Jones:2021wy}
{Jones}, M.~G., {Bennet}, P., {Mutlu-Pakdil}, B., {et~al.} 2021, \apj, 919, 72

\bibitem[{{Kado-Fong} {et~al.}(2020){Kado-Fong}, {Greene}, {Huang}, {Beaton}, {Goulding}, \& {Komiyama}}]{Kado-Fong:2020wg}
{Kado-Fong}, E., {Greene}, J.~E., {Huang}, S., {et~al.} 2020, \apj, 900, 163

\bibitem[{{Kado-Fong} {et~al.}(2022){Kado-Fong}, {Greene}, {Huang}, \& {Goulding}}]{kado-fong:2022_gas_rich_udgs}
{Kado-Fong}, E., {Greene}, J.~E., {Huang}, S., \& {Goulding}, A. 2022, \apj, 941, 11, \dodoi{10.3847/1538-4357/ac9964}

\bibitem[{{Kauffmann} {et~al.}(1993){Kauffmann}, {White}, \& {Guiderdoni}}]{Kauffmann:1993aa}
{Kauffmann}, G., {White}, S.~D.~M., \& {Guiderdoni}, B. 1993, \mnras, 264, 201

\bibitem[{{Koda} {et~al.}(2015){Koda}, {Yagi}, {Yamanoi}, \& {Komiyama}}]{Koda_2015_UDGs}
{Koda}, J., {Yagi}, M., {Yamanoi}, H., \& {Komiyama}, Y. 2015, \apjl, 807, L2, \dodoi{10.1088/2041-8205/807/1/L2}

\bibitem[{{Lacey} \& {Cole}(1993)}]{Lacey_Cole_1993}
{Lacey}, C., \& {Cole}, S. 1993, \mnras, 262, 627, \dodoi{10.1093/mnras/262.3.627}

\bibitem[{{Lacey} {et~al.}(2016){Lacey}, {Baugh}, {Frenk}, {Benson}, {Bower}, {Cole}, {Gonzalez-Perez}, {Helly}, {Lagos}, \& {Mitchell}}]{Lacey:2016aa}
{Lacey}, C.~G., {Baugh}, C.~M., {Frenk}, C.~S., {et~al.} 2016, \mnras, 462, 3854

\bibitem[{{Le Bret} {et~al.}(2015){Le Bret}, {Pontzen}, {Cooper}, {Frenk}, {Zolotov}, {Brooks}, {Governato}, \& {Parry}}]{Le-Bret:2015aa}
{Le Bret}, T., {Pontzen}, A., {Cooper}, A.~P., {et~al.} 2015, ArXiv e-prints

\bibitem[{{Le Bret} {et~al.}(2017){Le Bret}, {Pontzen}, {Cooper}, {Frenk}, {Zolotov}, {Brooks}, {Governato}, \& {Parry}}]{Cooper:2017aa}
---. 2017, \mnras, 468, 3212, \dodoi{10.1093/mnras/stx552}

\bibitem[{{Li} {et~al.}(2010){Li}, {De Lucia}, \& {Helmi}}]{Li:2010aa}
{Li}, Y.-S., {De Lucia}, G., \& {Helmi}, A. 2010, \mnras, 401, 2036

\bibitem[{{Lynden-Bell} \& {Lynden-Bell}(1995)}]{Lynden-Bell:1995vu}
{Lynden-Bell}, D., \& {Lynden-Bell}, R.~M. 1995, \mnras, 275, 429

\bibitem[{{Martin} {et~al.}(2013){Martin}, {Slater}, {Schlafly}, {Morganson}, {Rix}, {Bell}, {Laevens}, {Bernard}, {Ferguson}, {Finkbeiner}, {Burgett}, {Chambers}, {Hodapp}, {Kaiser}, {Kudritzki}, {Magnier}, {Morgan}, {Price}, {Tonry}, \& {Wainscoat}}]{martin:2013_andxxxii}
{Martin}, N.~F., {Slater}, C.~T., {Schlafly}, E.~F., {et~al.} 2013, \apj, 772, 15, \dodoi{10.1088/0004-637X/772/1/15}

\bibitem[{{McConnachie}(2012)}]{McConnachie_local_group_2021}
{McConnachie}, A.~W. 2012, \aj, 144, 4

\bibitem[{{Montes} {et~al.}(2020){Montes}, {Infante-Sainz}, {Madrigal-Aguado}, {Rom{\'a}n}, {Monelli}, {Borlaff}, \& {Trujillo}}]{Montes:2020aa_df4}
{Montes}, M., {Infante-Sainz}, R., {Madrigal-Aguado}, A., {et~al.} 2020, \apj, 904, 114

\bibitem[{{Mu{\~n}oz} {et~al.}(2015){Mu{\~n}oz}, {Eigenthaler}, {Puzia}, {Taylor}, {Ordenes-Brice{\~n}o}, {Alamo-Mart{\'{\i}}nez}, {Ribbeck}, {{\'A}ngel}, {Capaccioli}, {C{\^o}t{\'e}}, {Ferrarese}, {Galaz}, {Hempel}, {Hilker}, {Jord{\'a}n}, {Lan{\c c}on}, {Mieske}, {Paolillo}, {Richtler}, {S{\'a}nchez-Janssen}, \& {Zhang}}]{Munoz:2015aa}
{Mu{\~n}oz}, R.~P., {Eigenthaler}, P., {Puzia}, T.~H., {et~al.} 2015, \apjl, 813, L15

\bibitem[{{Newton} {et~al.}(2023){Newton}, {Di Cintio}, {Cardona-Barrero}, {Libeskind}, {Hoffman}, {Knebe}, {Sorce}, {Steinmetz}, \& {Tempel}}]{Newton:2023_hestia_local_group}
{Newton}, O., {Di Cintio}, A., {Cardona-Barrero}, S., {et~al.} 2023, \apjl, 946, L37, \dodoi{10.3847/2041-8213/acc2bb}

\bibitem[{{Okamoto} {et~al.}(2008){Okamoto}, {Gao}, \& {Theuns}}]{OkamotoGaoTheuns}
{Okamoto}, T., {Gao}, L., \& {Theuns}, T. 2008, \mnras, 390, 920, \dodoi{10.1111/j.1365-2966.2008.13830.x}

\bibitem[{{Pandya} {et~al.}(2018){Pandya}, {Romanowsky}, {Laine}, {Brodie}, {Johnson}, {Glaccum}, {Villaume}, {Cuillandre}, {Gwyn}, {Krick}, {Lasker}, {Mart{\'\i}n-Navarro}, {Martinez-Delgado}, \& {van Dokkum}}]{pandya2018}
{Pandya}, V., {Romanowsky}, A.~J., {Laine}, S., {et~al.} 2018, \apj, 858, 29, \dodoi{10.3847/1538-4357/aab498}

\bibitem[{Parkinson {et~al.}(2007)Parkinson, Cole, \& Helly}]{Parkinson2007}
Parkinson, H., Cole, S., \& Helly, J. 2007, Monthly Notices of the Royal Astronomical Society, 383, 557, \dodoi{10.1111/j.1365-2966.2007.12517.x}

\bibitem[{{Peng} \& {Lim}(2016)}]{peng2016}
{Peng}, E.~W., \& {Lim}, S. 2016, \apjl, 822, L31, \dodoi{10.3847/2041-8205/822/2/L31}

\bibitem[{{Pereira-Wilson} {et~al.}(2023){Pereira-Wilson}, {Navarro}, {Ben{\'\i}tez-Llambay}, \& {Santos-Santos}}]{Pereira-Wilson:2023aa}
{Pereira-Wilson}, M., {Navarro}, J.~F., {Ben{\'\i}tez-Llambay}, A., \& {Santos-Santos}, I. 2023, \mnras, 519, 1425

\bibitem[{{Planck Collaboration} {et~al.}(2016){Planck Collaboration}, {Adam}, {Aghanim}, {Ashdown}, {Aumont}, {Baccigalupi}, {Ballardini}, {Banday}, {Barreiro}, {Bartolo}, {Basak}, {Battye}, {Benabed}, {Bernard}, {Bersanelli}, {Bielewicz}, {Bock}, {Bonaldi}, {Bonavera}, {Bond}, {Borrill}, {Bouchet}, {Boulanger}, {Bucher}, {Burigana}, {Calabrese}, {Cardoso}, {Carron}, {Chiang}, {Colombo}, {Combet}, {Comis}, {Couchot}, {Coulais}, {Crill}, {Curto}, {Cuttaia}, {Davis}, {de Bernardis}, {de Rosa}, {de Zotti}, {Delabrouille}, {Di Valentino}, {Dickinson}, {Diego}, {Dor{\'e}}, {Douspis}, {Ducout}, {Dupac}, {Elsner}, {En{\ss}lin}, {Eriksen}, {Falgarone}, {Fantaye}, {Finelli}, {Forastieri}, {Frailis}, {Fraisse}, {Franceschi}, {Frolov}, {Galeotta}, {Galli}, {Ganga}, {G{\'e}nova-Santos}, {Gerbino}, {Ghosh}, {Gonz{\'a}lez-Nuevo}, {G{\'o}rski}, {Gruppuso}, {Gudmundsson}, {Hansen}, {Helou}, {Henrot-Versill{\'e}}, {Herranz}, {Hivon}, {Huang}, {Ili{\'c}}, {Jaffe}, {Jones}, {Keih{\"a}nen}, {Keskitalo}, {Kisner}, {Knox},
  {Krachmalnicoff}, {Kunz}, {Kurki-Suonio}, {Lagache}, {L{\"a}hteenm{\"a}ki}, {Lamarre}, {Langer}, {Lasenby}, {Lattanzi}, {Lawrence}, {Le Jeune}, {Levrier}, {Lewis}, {Liguori}, {Lilje}, {L{\'o}pez-Caniego}, {Ma}, {Mac{\'\i}as-P{\'e}rez}, {Maggio}, {Mangilli}, {Maris}, {Martin}, {Mart{\'\i}nez-Gonz{\'a}lez}, {Matarrese}, {Mauri}, {McEwen}, {Meinhold}, {Melchiorri}, {Mennella}, {Migliaccio}, {Miville-Desch{\^e}nes}, {Molinari}, {Moneti}, {Montier}, {Morgante}, {Moss}, {Naselsky}, {Natoli}, {Oxborrow}, {Pagano}, {Paoletti}, {Partridge}, {Patanchon}, {Patrizii}, {Perdereau}, {Perotto}, {Pettorino}, {Piacentini}, {Plaszczynski}, {Polastri}, {Polenta}, {Puget}, {Rachen}, {Racine}, {Reinecke}, {Remazeilles}, {Renzi}, {Rocha}, {Rossetti}, {Roudier}, {Rubi{\~n}o-Mart{\'\i}n}, {Ruiz-Granados}, {Salvati}, {Sandri}, {Savelainen}, {Scott}, {Sirri}, {Sunyaev}, {Suur-Uski}, {Tauber}, {Tenti}, {Toffolatti}, {Tomasi}, {Tristram}, {Trombetti}, {Valiviita}, {Van Tent}, {Vielva}, {Villa}, {Vittorio}, {Wandelt}, {Wehus}, {White},
  {Zacchei}, \& {Zonca}}]{planck_reionization_2016}
{Planck Collaboration}, {Adam}, R., {Aghanim}, N., {et~al.} 2016, \aap, 596, A108, \dodoi{10.1051/0004-6361/201628897}

\bibitem[{{Rees}(1986)}]{Rees:1986aa}
{Rees}, M.~J. 1986, \mnras, 218, 25P

\bibitem[{{Ricotti} {et~al.}(2022){Ricotti}, {Polisensky}, \& {Cleland}}]{Ricotti:2022aa}
{Ricotti}, M., {Polisensky}, E., \& {Cleland}, E. 2022, \mnras, 515, 302, \dodoi{10.1093/mnras/stac1485}

\bibitem[{{Sales} {et~al.}(2020){Sales}, {Navarro}, {Pe{\~n}afiel}, {Peng}, {Lim}, \& {Hernquist}}]{Sales:2020vg}
{Sales}, L.~V., {Navarro}, J.~F., {Pe{\~n}afiel}, L., {et~al.} 2020, \mnras, 494, 1848

\bibitem[{{Sand} {et~al.}(2022){Sand}, {Mutlu-Pakdil}, {Jones}, {Karunakaran}, {Wang}, {Yang}, {Chiti}, {Bennet}, {Crnojevi{\'c}}, \& {Spekkens}}]{sand:2022aa_tucb}
{Sand}, D.~J., {Mutlu-Pakdil}, B., {Jones}, M.~G., {et~al.} 2022, \apjl, 935, L17, \dodoi{10.3847/2041-8213/ac85ee}

\bibitem[{Sawala {et~al.}(2015)Sawala, Frenk, Fattahi, Navarro, Bower, Crain, Vecchia, Furlong, Jenkins, McCarthy, Qu, Schaller, Schaye, \& Theuns}]{Sawala_2015}
Sawala, T., Frenk, C.~S., Fattahi, A., {et~al.} 2015, Monthly Notices of the Royal Astronomical Society, 448, 2941, \dodoi{10.1093/mnras/stu2753}

\bibitem[{Simon(2019)}]{UDF_Simon_2019}
Simon, J.~D. 2019, Annual Review of Astronomy and Astrophysics, 57, 375, \dodoi{10.1146/annurev-astro-091918-104453}

\bibitem[{{The Astropy Collaboration} {et~al.}(2013){The Astropy Collaboration}, Robitaille, Tollerud, Greenfield, Droettboom, Bray, Aldcroft, Davis, Ginsburg, Price-Whelan, Kerzendorf, Conley, Crighton, Barbary, Muna, Ferguson, Grollier, Parikh, Nair, Unther, Deil, Woillez, Conseil, Kramer, Turner, Singer, Fox, Weaver, Zabalza, Edwards, Azalee~Bostroem, Burke, Casey, Crawford, Dencheva, Ely, Jenness, Labrie, Lim, Pierfederici, Pontzen, Ptak, Refsdal, Servillat, \& Streicher}]{Collaboration2013Astropy:Astronomy}
{The Astropy Collaboration}, Robitaille, T.~P., Tollerud, E.~J., {et~al.} 2013, A{\&}A, 558, A33, \dodoi{10.1051/0004-6361/201322068}

\bibitem[{{The Astropy Collaboration} {et~al.}(2018){The Astropy Collaboration}, Price-Whelan, Sip{\H{o}}cz, G{\"{u}}nther, Lim, Crawford, Conseil, Shupe, Craig, Dencheva, Ginsburg, VanderPlas, Bradley, P{\'{e}}rez-Su{\'{a}}rez, de~Val-Borro, Aldcroft, Cruz, Robitaille, Tollerud, Ardelean, Babej, Bach, Bachetti, Bakanov, Bamford, Barentsen, Barmby, Baumbach, Berry, Biscani, Boquien, Bostroem, Bouma, Brammer, Bray, Breytenbach, Buddelmeijer, Burke, Calderone, Cano~Rodr{\'{i}}guez, Cara, Cardoso, Cheedella, Copin, Corrales, Crichton, D'Avella, Deil, Depagne, Dietrich, Donath, Droettboom, Earl, Erben, Fabbro, Ferreira, Finethy, Fox, Garrison, Gibbons, Goldstein, Gommers, Greco, Greenfield, Groener, Grollier, Hagen, Hirst, Homeier, Horton, Hosseinzadeh, Hu, Hunkeler, Ivezi{\'{c}}, Jain, Jenness, Kanarek, Kendrew, Kern, Kerzendorf, Khvalko, King, Kirkby, Kulkarni, Kumar, Lee, Lenz, Littlefair, Ma, Macleod, Mastropietro, McCully, Montagnac, Morris, Mueller, Mumford, Muna, Murphy, Nelson, Nguyen, Ninan, N{\"{o}}the,
  Ogaz, Oh, Parejko, Parley, Pascual, Patil, Patil, Plunkett, Prochaska, Rastogi, Reddy~Janga, Sabater, Sakurikar, Seifert, Sherbert, Sherwood-Taylor, Shih, Sick, Silbiger, Singanamalla, Singer, Sladen, Sooley, Sornarajah, Streicher, Teuben, Thomas, Tremblay, Turner, Terr{\'{o}}n, van Kerkwijk, de~la Vega, Watkins, Weaver, Whitmore, Woillez, Zabalza, \& Contributors}]{Collaboration2018ThePackage}
{The Astropy Collaboration}, Price-Whelan, A.~M., Sip{\H{o}}cz, B.~M., {et~al.} 2018, AJ, 156, 123, \dodoi{10.3847/1538-3881/AABC4F}

\bibitem[{{Thoul} \& {Weinberg}(1996)}]{Thoul:1996aa}
{Thoul}, A.~A., \& {Weinberg}, D.~H. 1996, \apj, 465, 608

\bibitem[{{Torrealba} {et~al.}(2016){Torrealba}, {Koposov}, {Belokurov}, \& {Irwin}}]{Torrealba:2016ab}
{Torrealba}, G., {Koposov}, S.~E., {Belokurov}, V., \& {Irwin}, M. 2016, \mnras, 459, 2370

\bibitem[{Torrealba {et~al.}(2019)Torrealba, Belokurov, Koposov, Li, Walker, Sanders, Geringer-Sameth, Zucker, Kuehn, Evans, \& Dehnen}]{Torrealba_2019_Antlia2}
Torrealba, G., Belokurov, V., Koposov, S.~E., {et~al.} 2019, Monthly Notices of the Royal Astronomical Society, 488, 2743, \dodoi{10.1093/mnras/stz1624}

\bibitem[{{Trentham} {et~al.}(2001){Trentham}, {Tully}, \& {Verheijen}}]{Trentham:2001aa}
{Trentham}, N., {Tully}, R.~B., \& {Verheijen}, M.~A.~W. 2001, \mnras, 325, 385

\bibitem[{{van der Burg} {et~al.}(2017){van der Burg}, {Hoekstra}, {Muzzin}, {Sif{\'o}n}, {Viola}, {Bremer}, {Brough}, {Driver}, {Erben}, {Heymans}, {Hildebrandt}, {Holwerda}, {Klaes}, {Kuijken}, {McGee}, {Nakajima}, {Napolitano}, {Norberg}, {Taylor}, \& {Valentijn}}]{van-der-Burg:2017aa}
{van der Burg}, R. F.~J., {Hoekstra}, H., {Muzzin}, A., {et~al.} 2017, \aap, 607, A79, \dodoi{10.1051/0004-6361/201731335}

\bibitem[{{van Dokkum} {et~al.}(2019){van Dokkum}, {Danieli}, {Abraham}, {Conroy}, \& {Romanowsky}}]{PvD2019_2ndMissingDM}
{van Dokkum}, P., {Danieli}, S., {Abraham}, R., {Conroy}, C., \& {Romanowsky}, A.~J. 2019, \apjl, 874, L5, \dodoi{10.3847/2041-8213/ab0d92}

\bibitem[{{van Dokkum} {et~al.}(2015){van Dokkum}, {Abraham}, {Merritt}, {Zhang}, {Geha}, \& {Conroy}}]{van_Dokkum_2015}
{van Dokkum}, P.~G., {Abraham}, R., {Merritt}, A., {et~al.} 2015, \apjl, 798, L45, \dodoi{10.1088/2041-8205/798/2/L45}

\bibitem[{{Van Nest} {et~al.}(2022){Van Nest}, {Munshi}, {Wright}, {Tremmel}, {Brooks}, {Nagai}, \& {Quinn}}]{Van-Nest:2022wg}
{Van Nest}, J.~D., {Munshi}, F., {Wright}, A.~C., {et~al.} 2022, \apj, 926, 92

\bibitem[{Wang {et~al.}(2021)Wang, Nadler, Mao, Adhikari, Wechsler, \& Behroozi}]{Wang2021}
Wang, Y., Nadler, E.~O., Mao, Y.-Y., {et~al.} 2021, The Astrophysical Journal, 915, 116, \dodoi{10.3847/1538-4357/ac024a}

\bibitem[{{Weinmann} {et~al.}(2007){Weinmann}, {Macci{\`o}}, {Iliev}, {Mellema}, \& {Moore}}]{Weinmann_2007}
{Weinmann}, S.~M., {Macci{\`o}}, A.~V., {Iliev}, I.~T., {Mellema}, G., \& {Moore}, B. 2007, \mnras, 381, 367, \dodoi{10.1111/j.1365-2966.2007.12279.x}

\bibitem[{{White} \& {Frenk}(1991)}]{White_Frenk_1991}
{White}, S. D.~M., \& {Frenk}, C.~S. 1991, \apj, 379, 52, \dodoi{10.1086/170483}

\bibitem[{{White} \& {Rees}(1978)}]{White:1978aa}
{White}, S.~D.~M., \& {Rees}, M.~J. 1978, \mnras, 183, 341

\bibitem[{{Wright} {et~al.}(2017){Wright}, {Robotham}, {Driver}, {Alpaslan}, {Andrews}, {Baldry}, {Bland-Hawthorn}, {Brough}, {Brown}, {Colless}, {da Cunha}, {Davies}, {Graham}, {Holwerda}, {Hopkins}, {Kafle}, {Kelvin}, {Loveday}, {Maddox}, {Meyer}, {Moffett}, {Norberg}, {Phillipps}, {Rowlands}, {Taylor}, {Wang}, \& {Wilkins}}]{Wright:2017aa}
{Wright}, A.~H., {Robotham}, A.~S.~G., {Driver}, S.~P., {et~al.} 2017, \mnras, 470, 283

\bibitem[{{Zaritsky} {et~al.}(2023){Zaritsky}, {Donnerstein}, {Dey}, {Karunakaran}, {Kadowaki}, {Khim}, {Spekkens}, \& {Zhang}}]{zaritsky:2023_smudges}
{Zaritsky}, D., {Donnerstein}, R., {Dey}, A., {et~al.} 2023, \apjs, 267, 27, \dodoi{10.3847/1538-4365/acdd71}

\end{thebibliography}
\bibliographystyle{aasjournal}



\end{document}